\documentclass[10pt]{article}

\usepackage{octen_report}
\usepackage[utf8]{inputenc}
\usepackage{microtype}
\usepackage{amsmath,amssymb}
\usepackage{booktabs}
\usepackage{float}          
\usepackage{flafter}        
\usepackage{array}
\usepackage{multirow}
\usepackage{graphicx}
\usepackage{xcolor}
\usepackage{enumitem}
\usepackage{fontawesome5}   
\usepackage[colorlinks=true,linkcolor=blue!50!black,citecolor=blue!50!black,urlcolor=blue!50!black]{hyperref}
\usepackage[capitalise,noabbrev]{cleveref}
\hypersetup{pdftitle={Octen-WebGR: Scaling Generative Retrieval to 200 Million Web Pages},pdfauthor={Guangwei Xu, Wenkai Zhang, Yawen Liu, Dongxu Zhang, Xijie Xu, Kuan Zou}}

\newcommand{\model}{Octen-WebGR-4B}
\newcommand{\modelsmall}{Octen-WebGR-0.6B}
\newcommand{\family}{Octen-WebGR}
\newcommand{\Ucore}{$\mathcal{U}_{\text{100M}}$}
\newcommand{\Ufull}{$\mathcal{U}_{\text{201M}}$}
\newcommand{\Usmall}{$\mathcal{U}_{\text{31M}}$}
\newcommand{\trafficsep}{\textsc{Traffic-Sep}}     
\newcommand{\trafficjul}{\textsc{Traffic-Jul}}     
\newcommand{\heldnav}{\textsc{Held-Nav}}

\newcommand{\twoline}[2]{\begin{tabular}[b]{@{}c@{}}#1\\#2\end{tabular}}

\AtBeginDocument{\setlength{\parfillskip}{0pt plus .75\textwidth}}
\DeclareCaptionJustification{noshortlast}{\setlength{\parfillskip}{0pt plus .75\linewidth}}
\title{\family: Scaling Generative Retrieval\\to 200 Million Web Pages}
\author{Guangwei Xu, Wenkai Zhang, Yawen Liu, Dongxu Zhang, Xijie Xu, Kuan Zou}
\reportemail{research@octen.ai}
\reportlink{\includegraphics[height=9pt]{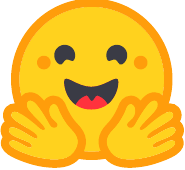}}{https://huggingface.co/Octen}
\reportlink{\octenicon}{https://octen.ai}
\reportlink{\faGithub}{https://github.com/Octen-Team}

\begin{document}
\maketitle

\begin{abstract}
Navigational search aims to identify a single target page from hundreds of millions of web pages. Existing approaches face a
fundamental trade-off: click tables perform well for queries seen before but cannot generalise to new query formulations, while
embedding-based retrievers can generalise across content but often struggle to rank navigational targets accurately. To address this
challenge, we introduce \family, a generative retrieval framework that stores web page representations directly in the parameters of a
language model. Given a query, the model generates a compact two-token semantic identifier that is constrained to a table of real URLs,
enabling direct retrieval of a specific web page. We develop \model, which indexes 201 million web pages, and evaluate it on real-world
navigational search traffic from the month following training. Our results show that generative retrieval is particularly effective at handling new ways of expressing requests for familiar pages.
For unseen queries targeting pages with substantial historical interaction data, \family{} ranks the correct page first 30.9\% of the
time, compared with 17.2\% for a dense retriever trained on the same click data (21.4\% with a click prior). Conversely, embedding-based methods retain an advantage
on rarely visited pages, highlighting the complementary strengths of query-side generalisation in generative retrieval and page-side
generalisation in embedding retrieval. Across overall navigational traffic, \family{} achieves a top-1 accuracy of 59.6\%, substantially
outperforming a click-enhanced dense retriever (46.5\%) and frontier large language models answering from memory (up to 37.5\%), which
cost at least 40 times as much per query. Beyond effectiveness, \family{} is practical at web scale. The 4B-parameter model operates with a single GPU and 10\,GB of memory,
compared with 412\,GB required for exact dense retrieval. Scaling experiments with up to 32 million URLs show that, when every query--page pair is seen equally often, the model size needed grows sublinearly with index size, and that retrieval performance is constrained more by page exposure than model capacity. We release both \model{}
and \modelsmall, together with their identifier tables and inference code, to support further research on web-scale generative
retrieval.
\end{abstract}

\begin{figure}[!h]
  \centering
  \includegraphics[width=\linewidth]{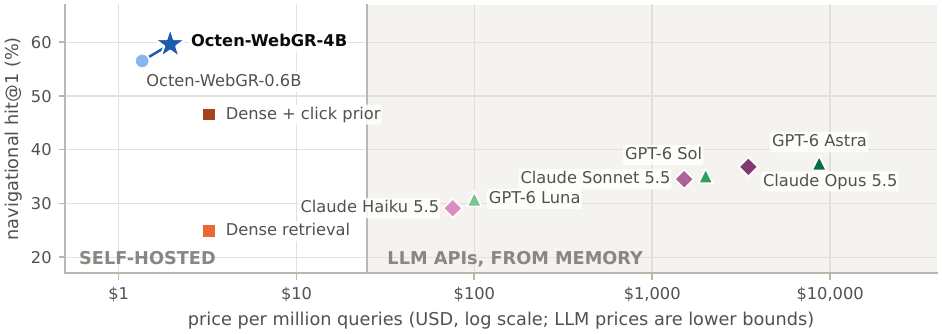}
  \caption{\textbf{Accuracy against price.} Navigational hit@1 on September traffic, the share of navigational traffic whose target
  page a system ranks first, against the price of answering one million queries. Self-hosted systems are priced at \$3 per H200 GPU-hour of measured compute, leaving out the 412\,GB index that exact dense search keeps in memory. LLMs answer from memory at API list prices, a lower bound, and are scored on a 5,000-query sample, on
  which \model{} reaches 59.8\%. Details in \cref{tab:serving,tab:llm}.}
  \label{fig:cost}
\end{figure}

\section{Introduction}
\label{sec:intro}

Many web searches name a destination rather than ask a question. A user who types ``nyt connections'' wants one particular page,
the puzzle at \texttt{nytimes.com/games/connections}, and the search engine serves the query well only if it puts that page first.
\citet{broder2002taxonomy} called such queries \emph{navigational}; they were a fifth to a quarter of the queries he analysed. He set
them apart from \emph{informational} queries, which ask for content on a topic, and from \emph{transactional} queries, which seek a
service such as a purchase or a download. Since a service is reached through a page, we count a transactional query as navigational
whenever its clicks concentrate on one page (\cref{sec:task}). Navigational search is thus the task of finding one page among the
hundreds of millions that a search engine can return, and of ranking it first.

Search engines answer such queries with a click table or a dense retriever, and each covers only part of the traffic. A click table
returns the page that past users clicked for the same query, so it has no answer for a query it has never seen; in the traffic we study,
a third of navigational clicks come from such queries (\cref{sec:evaldata}). A dense retriever answers any query by comparing its
embedding with those of the pages (\cref{sec:background}), but similarity is not what navigation asks for. For almost a third of
navigational traffic, dense retrieval has the target among its first ten results but ranks another URL above it, often another page of
the same site (\cref{sec:mainres}). Large language models (LLMs)
asked for the URL from memory are neither accurate nor cheap enough (\cref{sec:llmres}), and they can name URLs that do not
exist~\citep{ziems2023llmurl,rao2026refhalluc}.

Generative retrieval (GR) takes a different route~\citep{metzler2021rethinking,tay2022dsi}: a language model is trained to write the
identifier of the page a query leads to, so that the model itself is the index. Navigation suits this design. Each navigational
query has one right answer, the answers come from a fixed list of URLs, and past clicks show which page each query led to. From these
examples a model can learn which of many similar pages users want, and its output can be restricted to URLs that exist.

GR has not yet been shown to work at web scale, however. It has reached the 8.8M passages of MS MARCO, and going beyond was called
``an important and unsolved challenge''~\citep{pradeep2023scale}. Larger corpora have remained difficult~\citep{zhang2023mevi,tang2024gr2},
and recent studies of how GR scales and where it fails still work at that scale or
below~\citep{cai2025scaling,li2025genirsurvey,bracher2026limit}. Generative systems with 100 million to a billion targets exist, but they index
videos, products, advertising keywords or whole sites~\citep{chen2025unisearch,ling2026crid,lian2019baidu,tang2023sedsi,lee2026authgr},
not individual web pages. Web pages raise three problems at once. Every one of hundreds of millions of pages needs a short identifier that names it and no other page (\cref{sec:sid}). The model must learn about a billion query--URL pairs of very uneven popularity over tens of
thousands of training steps, during which pairs learned early can be forgotten (\cref{sec:limits}). And many pages have been reached by only a few
queries, so the model has little to learn them from (\cref{sec:newpages}).

This report presents \family, to our knowledge the first generative retrievers over hundreds of millions of individual web pages. Each URL is
named by a two-token semantic identifier: two numbers between 0 and 65,535, assigned so that similar pages tend to share the first and
no two URLs share both. The model is trained on query--URL pairs from a proprietary click dataset to generate the identifier of the
clicked page, and a table maps a generated identifier back to its URL. We train on clicks up to August 2026 and evaluate forward in
time, on navigational traffic recorded in September. We release two models with their identifier tables and inference code: \model,
Qwen3-4B-Base fine-tuned to index 201,231,792 URLs, and \modelsmall, Qwen3-0.6B-Base fine-tuned to index 31.3M of them (every homepage
and the most-clicked other pages). Both run on a single GPU, and the smaller one also on one CPU socket (\cref{sec:serving}).

We measure accuracy mainly by hit@1, the share of navigational traffic whose target a system ranks first. Our main findings are three.
\begin{enumerate}[leftmargin=1.5em,itemsep=2pt,topsep=2pt]
  \item \textbf{GR is the most accurate system we measured, and it holds up as the number of URLs grows.} \model{} ranks the target
  first for 59.6\% of navigational traffic, against 46.5\% for a click-enhanced dense retriever and at most 37.5\% for frontier LLMs
  (\cref{sec:mainres,sec:llmres}). Going from 2 million to 201 million URLs costs it 4.7 points of hit@1 and dense retrieval
  22.2 (\cref{sec:ladder}).
  \item \textbf{GR generalises to new ways of asking for pages it knows; embeddings generalise to pages with little click history.} On
  new queries for pages with many training queries, GR ranks the target first for 30.9\%, against 17.2\% for a dense retriever trained on the same click data and 21.4\% with a click prior. On rarely clicked pages the order reverses: the two are complementary
  (\crefrange{sec:mainres}{sec:combo}).
  \item \textbf{GR is limited by exposure, not by capacity.} Up to 32 million URLs, what a model retains of a pair depends on
  how often, and how late in training, it saw the pair, not on how many other pairs it holds (\cref{sec:limits}). Scaling GR is
  therefore as much a matter of scheduling training as of model size.
\end{enumerate}
\Cref{sec:background} introduces the preliminaries and \crefrange{sec:taskdata}{sec:training} describe the task, the data and the
system. \Crefrange{sec:results}{sec:limits} report our experiments: how \family{} compares with other systems, how it scales and what
limits it. \Cref{sec:further} draws the consequences for larger universes, and \crefrange{sec:release}{sec:limitations} cover the release,
related work and limitations.

\section{Preliminaries}
\label{sec:background}

This section explains how dense and generative retrievers work and introduces the terms used throughout this report. We call the items returned
by a retriever documents; in our setting, these documents are web pages identified by their URLs.

\paragraph{Dense retrieval.} A dense retriever, or embedding retriever, uses a neural encoder to map queries and documents to vectors.
The encoder is trained so that a query lands close to the documents relevant to it~\citep{karpukhin2020dpr}, and the retriever answers
a query with the documents whose vectors are most similar to the query's, usually by cosine similarity. Because it compares meanings
rather than words, it generalises to new queries, and it is the standard tool for informational search~\citep{thakur2021beir,muennighoff2023mteb}.
It needs an index that stores a vector for every document. Searching that index exactly costs memory and time in proportion to the
number of documents, and compressed approximate indexes trade some accuracy for size~\citep{jegou2011pq,jayaramsubramanya2019diskann,chen2021spann}.
A larger index also holds more documents that can come close to a query by chance. \citet{reimers2021curse} found that, as the index
grows, the accuracy of dense retrievers falls faster than that of sparse retrievers, which match words, and the more so the fewer
dimensions the vectors have. \Cref{sec:ladder} measures this fall for our dense baseline and for GR.

\paragraph{Generative retrieval.} A generative retriever produces the identifier of a document one token at a time, as a language model
produces text~\citep{metzler2021rethinking,tay2022dsi}. It learns the corpus from training samples that map queries to the identifiers
of their relevant documents, often together with \emph{indexing} samples that map a document's own text to its
identifier~\citep{tay2022dsi}. Because the corpus lives in the weights, learning new documents can erode what the model knew about old
ones~\citep{mehta2023dsipp}. \Cref{sec:limits} measures what governs this forgetting.

\paragraph{Semantic identifiers.} An identifier can be an arbitrary number, a string such as a title or a
URL~\citep{decao2021genre,zhou2022ultron}, or a \emph{semantic identifier} (SID), a short sequence of codes in which similar documents
share a prefix~\citep{tay2022dsi,rajput2023tiger}. SIDs are usually built by residual quantisation~\citep{chen2010rvq}. Each document is
embedded as a vector, and $k$-means clustering assigns the vector to its nearest centroid, whose index is the first code. A second
$k$-means clusters the \emph{residuals}, the vectors minus their centroids, and gives the second code. Documents that are close in the
embedding space thus share their first code and are told apart by the second. Nothing in this procedure keeps two documents from
receiving the same codes, however, whereas a retriever needs every identifier to name exactly one document; \cref{sec:sid} describes
how \family{} ensures this.

\paragraph{Constrained decoding.} GR produces its results by beam search, which keeps the $k$ most probable partial identifiers at
each step and returns $k$ complete ones, ranked by probability. Decoding is constrained by a prefix tree, or \emph{trie}, of all valid
identifiers: at each step, only tokens that continue some valid identifier are allowed~\citep{decao2021genre}. Every output therefore
names an existing document, and deleting an identifier from the trie removes its document from every possible output.

\section{Task and Data}
\label{sec:taskdata}

This section defines the task and its metric, and describes the click data and the URL universes.

\begin{table}[t]
  \centering
  \small
  \caption{\textbf{The three URL universes} (\cref{sec:universes}). Click share is the universe's share of the pool's clicks;
  coverage is the share of September navigational traffic whose target is in the universe.}
  \label{tab:universe}
  \begin{tabular}{lrrr}
    \toprule
     & \Usmall & \Ucore & \Ufull \\
    \midrule
    URLs & 31,282,165 & 100,478,486 & 201,231,792 \\
    \quad homepages & 15,535,349 & 15,535,349 & 15,535,349 \\
    \quad channel pages & 4,957,423 & 38,234,014 & 43,358,161 \\
    \quad deeper pages & 10,789,393 & 46,709,123 & 142,338,282 \\
    Click share of the pool & 74.7\% & 86.5\% & 92.5\% \\
    Coverage of \trafficsep{} navigational traffic & 68.9\% & 74.9\% & 79.2\% \\
    \bottomrule
  \end{tabular}
\end{table}

\subsection{Task definition}
\label{sec:task}

A \emph{universe} $\mathcal{U}$ is the set of URLs a retriever may return. For a query $q$ issued in an evaluation month, the
\emph{target} $u^*(q)$ is the URL that received the most clicks for $q$ in that month. Following \citet{teevan2011pnav}, who identify
navigational queries by the concentration of their clicks on a single result, we call $q$ \emph{navigational} when its target received
at least 10 clicks and at least half of the clicks on $q$. Our metric is navigational hit@$k$, the probability that the target is among
the top $k$ results for a randomly drawn navigational click. For a sample $C$ of navigational clicks, where click $c$ was made on query
$q_c$ and $R_k(q)$ denotes a system's top $k$ results,
\begin{equation}
  \text{hit@}k \;=\; \frac{1}{|C|} \sum_{c \in C} \mathbf{1}\left[\, u^*(q_c) \in R_k(q_c) \,\right].
  \label{eq:hitk}
\end{equation}
Because clicks rather than queries are drawn, each query counts in proportion to its traffic, and a target outside a system's universe
counts as a miss. We report hit@1 and hit@10. A navigational query has one right answer, so hit@1 is the measure that matters in web
search; hit@10 shows whether the target is at least within reach of a reranker.

\subsection{Click data}
\label{sec:clickdata}

Our training signal is a proprietary query--URL click dataset that Octen built and maintains in developing its web search API. It
records aggregated click counts between normalised queries and URLs and carries no user identifiers. We train on data up to the end of
August 2026, use July traffic for development and evaluate on traffic recorded in September (\cref{sec:evaldata}). Queries are normalised by collapsing whitespace and lower-casing; URLs by lower-casing and removing the scheme, a
leading \texttt{www.} and a trailing slash. \Cref{fig:overview}a shows three rows of the data: the queries ``connections'', ``nyt
connections'' and ``connections game'' send 74\%, 66\% and 49\% of their clicks to \texttt{nytimes.com/games/connections}, and the rest
of their clicks to other URLs.

\subsection{URL universes}
\label{sec:universes}

A model can return only the URLs of its universe, so the universe caps its accuracy. We use three nested universes
(\cref{tab:universe}). The main one, \Ufull{} (201,231,792 URLs), is drawn from a pool of 349M clicked URLs in two parts. \Ucore{}
(100.5M URLs over 18.9M hosts) keeps every homepage and the channel and deeper pages above click thresholds. To it we add 100.8M
\emph{navigational destination pages}, the dominant click targets of navigational queries. We call the latter \emph{new pages} and the
URLs of \Ucore{} \emph{old pages}. Adding the new pages raised the coverage of September navigational traffic from 74.9\% for \Ucore{} to
79.2\% for \Ufull, the most that any system returning URLs from \Ufull{} can reach. \Usmall{} (31.3M URLs), the universe of \modelsmall,
keeps every homepage and the most-clicked other pages. Restricted to \Ufull, the training data hold 1.13 billion query--URL pairs over
479 million queries.

\section{Method}
\label{sec:method}

\begin{figure}[t]
  \centering
  \includegraphics[width=\linewidth]{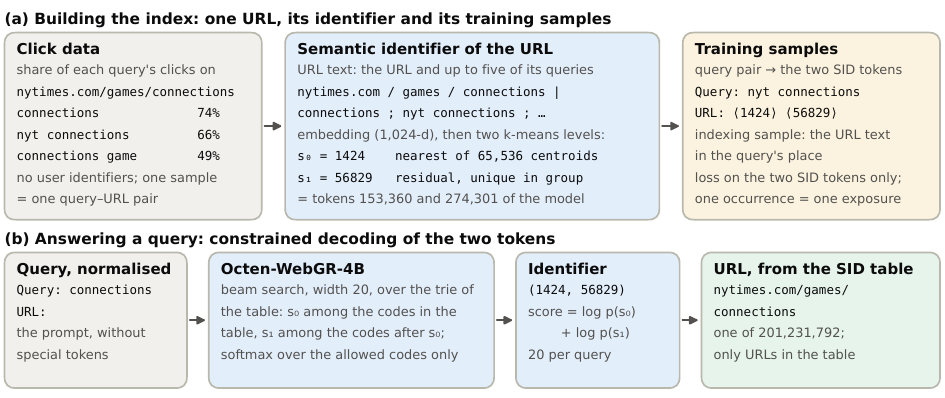}
  \caption{\textbf{\family{} on one real URL.} (a) How the URL receives its identifier and the samples the model is trained on
  (\cref{sec:sid,sec:samples}). (b) How a query is answered (\cref{sec:inference}). All strings are real, and the output in (b) is a
  real result on September traffic.}
  \label{fig:overview}
\end{figure}

\family{} models are decoder-only language models of the Qwen3 family~\citep{yang2025qwen3}, fine-tuned in full. Every URL is named by
a pair of new tokens added to the model's vocabulary, and the model learns to generate the pair from the queries that lead to the URL.
\Cref{fig:overview} follows one real URL through the system: how it receives its identifier (\cref{sec:sid}), the samples the model is
trained on (\cref{sec:samples}), and how a query is answered (\cref{sec:inference}). \Cref{sec:extend} explains how the identifiers were
extended when the universe grew.

\subsection{Semantic identifiers}
\label{sec:sid}

Every URL in a universe receives a two-token semantic identifier (SID) $(s_0, s_1)$. We call $s_0$ the first code and $s_1$ the second
code. Each is a number from 0 to 65,535, that is, an unsigned 16-bit integer. The URL of \cref{fig:overview}, \texttt{nytimes.com/\allowbreak games/\allowbreak connections},
has the SID $(1424, 56829)$. Two tokens keep decoding to two steps, and two codes of 65,536 values leave room for far more URLs than
\Ufull{} holds (\cref{sec:cost}).

\paragraph{Building the codes.} The codes come from quantising an embedding of the URL, so that URLs with similar embeddings share
their first code (\cref{sec:background}). The text embedded is the URL's \emph{URL text}: its normalised form followed by up to five of
its most-clicked queries, for example ``\texttt{nytimes.com / games / connections | connections ; nyt connections ; connections nyt ;
\ldots}''. A URL text contains no page content. BM25 and the dense retriever index the same texts (\cref{sec:baselines}). The
encoder is that of the dense retriever: a checkpoint of Octen-Embedding-4B~\citep{octen2025embedding}, whose embedding is truncated to
its first 1,024 dimensions and passed through a linear map trained on navigational query--URL pairs from our click data. These
1,024-dimensional vectors, one per URL of \Ucore{}, are the data points of two rounds of $k$-means clustering
with 65,536 centroids each~\citep{chen2010rvq,zhou2025onerectr}. The first code $s_0$ is the index of the nearest centroid of the first
round. The second round clusters the residuals, and the second code $s_1$ is the index of the nearest residual centroid, unless the
uniqueness rule below assigns another.

\paragraph{Unique identifiers.} Plain residual quantisation would give 51.5\% of the URLs of \Ucore{} a pair of codes shared with
another URL. We instead make second-level codes unique within each first-level group by a capacity-constrained assignment, in the manner
of collision-free identifier schemes~\citep{zhang2026faithful,zhang2025psi,ling2026crid}. The assignment runs in proposal rounds: every
unassigned URL proposes its next-nearest residual centroid, and when several URLs of one first-level group propose the same code, the
closest keeps it. Every URL ends with a unique pair; 63.5\% keep their nearest residual centroid and 96.2\% receive one of their eight
nearest. Each SID therefore names exactly one URL, and all metrics are computed over URLs.

\paragraph{Identifier tokens.} The model writes an SID as two new tokens, one for each code. The same number means different things in
the two positions: as $s_0$ it names a group of similar URLs, as $s_1$ one URL within a group. We therefore add two sets of 65,536 tokens
to the vocabulary of Qwen3, one set for $s_0$ and one for $s_1$. They are numbered after the 151,936 tokens of Qwen3, first the set for
$s_0$, then the set for $s_1$, so that in the example $s_0 = 1424$ is token $151{,}936 + 1{,}424 = 153{,}360$ and $s_1 = 56829$ is token
$151{,}936 + 65{,}536 + 56{,}829 = 274{,}301$. Each new token needs an embedding, the vector that represents it inside the model. We
start it at the average of the final-layer hidden states that the model computes for the prompts of queries leading to URLs with that
code, so that from the start the model favours the code after similar queries~\citep{loveland2026centroid}.

\subsection{Training samples and objective}
\label{sec:samples}

A training sample is one query--URL pair $(q, u)$, a query $q$ and a URL $u$ on which it was clicked. It is written as the prompt
\texttt{Query:~\{q\}\textbackslash nURL:} followed by the two SID tokens of $u$ (\cref{fig:overview}a). The pairs of the training data are
drawn with weight
\begin{equation}
  w(q,u) \;=\; \frac{n(q,u)}{N_q}\, N_q^{\beta}, \qquad \beta = 0.5,
\end{equation}
where $n(q,u)$ is the number of clicks on $u$ for $q$ and $N_q$ the total number of clicks for $q$. Within a query, URLs are thus
sampled in proportion to their clicks, and a query's share of the samples grows with the square root of its clicks, so that popular
queries are seen more often without crowding out the rest. Two further sample types are mixed in. A \emph{navigational floor} guarantees
the dominant pair of each navigational query a minimum number of samples, so that rarely searched destinations are not left out.
\emph{Document-indexing} samples are the indexing samples of \cref{sec:background}. They put a page's URL text (\cref{sec:sid}) in the
query's place of the prompt and map it to the page's SID. Since a URL text lists the page's most-clicked queries, one such sample shows
the model several of the page's query--URL pairs at once. Old pages received these samples in the round on \Ucore{}, and a random
quarter to half of the new pages in each later round (\cref{sec:rounds}). Queries are truncated to 32 tokens and URL texts to 64.

The loss is computed on the two SID tokens only. At each position the softmax is restricted to the codes that continue the gold prefix
in the table, as in decoding (\cref{sec:inference}).

\subsection{Inference}
\label{sec:inference}

\Cref{fig:overview}b shows how a query is answered. The query is normalised by collapsing whitespace to single spaces, trimming it and
lower-casing it, and truncated to 32 tokens; the prompt is that of \cref{sec:samples}, without special tokens. The model then emits the
two identifier tokens by beam search of width 20 over the trie of the identifier table~\citep{decao2021genre,lian2019baidu}: the first
step allows only first codes that occur in the table, and the second only the second codes that follow the chosen first code. The
softmax is taken over the allowed codes, and an identifier is scored by the sum of the log-probabilities of its two codes. A lookup in
the table maps each of the 20 identifiers to its URL, so every result is an existing URL.

\subsection{Extending the universe}
\label{sec:extend}

The codebooks and the codes above were built on \Ucore. When the universe was extended to \Ufull, existing codes were kept, so that
training could continue from the model trained on \Ucore{} (\cref{sec:rounds}). The new URLs cluster, up to 320K of them on one first-level centroid, more than the 65,536 second-level codes of a group can tell apart. When extending, we therefore capped every first-level group at 16,384 URLs, old and new together, a quarter of what its codes allow, so that a crowded group spills into the next-nearest one. No group reaches the cap with the URLs of \Ucore{} alone, so the cap only limits how many new URLs a group takes; 91.9\% of the new URLs still receive their nearest first-level centroid. \Usmall{} keeps the SIDs of \Ufull, so a model trained on either universe can decode over the other.

\section{Training}
\label{sec:training}

This section describes how the released models were trained. It also defines exposures, the decay window and the cumulative learning
rate, on which the analysis of \cref{sec:limits} rests.

\subsection{Optimisation}
\label{sec:objective}

We train with AdamW ($\beta=(0.9, 0.95)$, no weight decay, gradients clipped at norm 1) and a warmup--stable--decay (WSD)
schedule~\citep{hu2024minicpm}: the learning rate rises to a peak of $10^{-4}$, stays there and decays at the end of a run. Training uses
fp32 master weights with bf16 autocast and FlashAttention-2. Samples are packed into rows of 16,384 tokens, two rows per GPU on 8 H200
GPUs. A step therefore holds 262,144 tokens, about 16,000 samples, and a step of the 4B model takes 3.1 seconds. Within a round the order
of samples is a random permutation, so where a query--URL pair falls and how often it is drawn are independent of how hard it is.

\subsection{Exposures and the decay window}
\label{sec:exposure}

\begin{figure}[t]
  \centering
  \includegraphics[width=\linewidth]{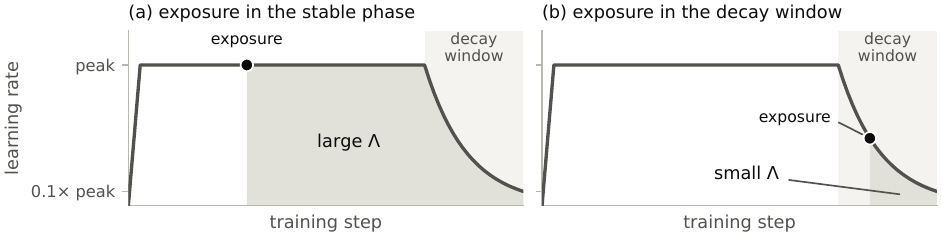}
  \caption{\textbf{Exposures and the cumulative learning rate}, schematically. The learning rate warms up, stays at its peak and decays
  over the final decay window (light band). The cumulative learning rate $\Lambda$ after an exposure is the area under the curve that
  follows it (grey): large for an exposure in the stable phase (a), small for one inside the decay window (b).}
  \label{fig:schedule}
\end{figure}

We call one occurrence of a query--URL pair in training an \emph{exposure}. A run ends with a \emph{decay window}, the final stretch over which the
learning rate decays (\cref{fig:schedule}). Two properties of an exposure matter later in the report: the learning rate at which it is
made, and how much training follows it. For an exposure at step $t$ of a run of $T$ steps, with learning rate $\eta_s$ at step $s$, the
\emph{cumulative learning rate} after the exposure is
\begin{equation}
  \Lambda \;=\; \sum_{s=t+1}^{T} \eta_s ,
  \label{eq:cumlr}
\end{equation}
the area under the learning-rate curve after the exposure. It is large for an exposure early in a run and small for one inside the decay
window. \Cref{sec:limits} shows that $\Lambda$ governs forgetting.

\subsection{Training rounds of \model}
\label{sec:rounds}

\model{} is trained in stages (\cref{tab:rounds} in \cref{app:window}). It starts from Qwen3-4B-Base with URL-string pre-training, in which the model learns to generate URLs as plain text (34K steps, 8.9B tokens). SID training follows: one round on \Ucore{} and four rounds on \Ufull, each initialised from the
previous end point. SID training totals 147.6K steps, 39B tokens and 1,019 H200 GPU-hours, or 127 hours on 8 GPUs. Each round re-warms
to the peak learning rate over 300 steps and decays to $0.1\times$ the peak. The decay windows of rounds 3 and 4 are curated: they hold
the dominant pairs of frequent navigational queries and the end of the main sample sequence, so that the pairs that carry the most
traffic are exposed where little training follows (\cref{sec:window}).

\subsection{Learning-rate schedule}
\label{sec:schedule}

Re-warming at the start of a round costs accuracy before it helps. Each round re-warmed to $10^{-4}$, and each time the hit@1 of the
previous end point on \trafficjul, our development set of July traffic (\cref{sec:evaldata}), dropped before recovering: by 4.6 points at
the start of round 2 and by 3.6 at the start of round 3~\citep{gupta2023rewarm,ibrahim2024simple}. Round 4 decays exponentially from the
peak to $0.1\times$ over its last 15,047 steps. From its checkpoint at step 6,000 we also ran a constant-rate arm ($3.5\times10^{-5}$, then
a linear decay over the last 2,257 steps). It serves as a control in \cref{sec:window} and ends 0.34 points of hit@1 below the released
arm.

\subsection{Training \modelsmall}
\label{sec:smalltrain}

\modelsmall{} is trained from Qwen3-0.6B-Base on \Usmall{} alone, with a simpler recipe that we also use for the 6,000-step comparisons of
\cref{sec:s1}: no URL-string pre-training, no document-indexing samples, a peak learning rate of $10^{-4}$ after 300 warm-up steps, and a linear decay
to $0.1\times$ over the final 10\% of steps. It runs for 43.9K steps, four passes over its samples, in 8.9 hours on 8 H200 GPUs.
\Cref{sec:smalluniverse} shows why a model of this size gains from the smaller universe.

\section{Evaluation on Search Traffic}
\label{sec:results}

This section and the next two report our experiments. This section compares \family{} with other systems on search traffic and asks
where its accuracy comes from. \Cref{sec:scaling} varies the number of URLs and the size of the model, and \cref{sec:limits} asks what
limits a model of a given size. Below, \cref{sec:evaldata} describes the evaluation data and \cref{sec:baselines} the systems we compare.

\subsection{Evaluation data}
\label{sec:evaldata}

\begin{figure}[t]
  \centering
  \includegraphics[width=\linewidth]{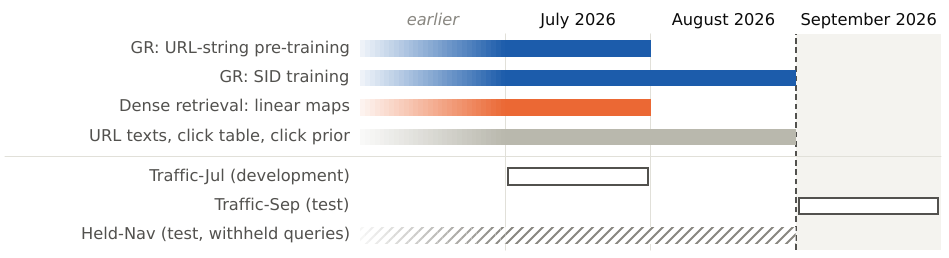}
  \caption{\textbf{The months behind each system and each evaluation set.} Bars show the click data that each component was trained
  or built on, which extend back before July (faded), and the months of the evaluation sets. The dashed line marks the end of the
  training data. URL texts are what BM25 indexes, what the dense retriever encodes and what the SIDs are built from. \heldnav{} holds queries from the training data that were withheld from SID
  training and from the click table, minus those that occur in the July data behind GR's URL-string pre-training and the dense
  retriever's maps. Closed-book LLMs use no click data.}
  \label{fig:timeline}
\end{figure}

\begin{figure}[t]
  \centering
  \includegraphics[width=\linewidth]{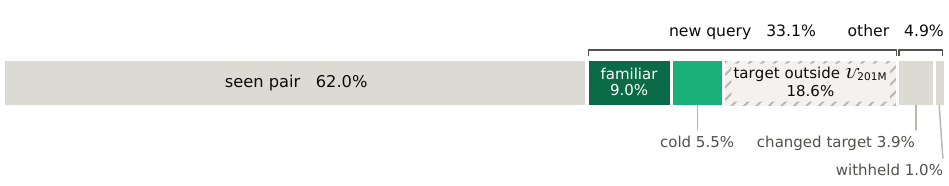}
  \caption{\textbf{Navigational traffic of \trafficsep{} by what a click table can know}, as shares of navigational clicks. The two
  coloured blocks, new queries whose target is in \Ufull, are where generalisation is tested; no system can answer the hatched block.}
  \label{fig:blocks}
\end{figure}

\Cref{fig:timeline} shows which months of data each evaluation set and each system draws on. \trafficsep{} is our main
evaluation set. It contains only clicks from September 2026, the month after the training data end, so every
model is tested on traffic that arrived after its training. It samples queries that received at least 10 clicks in September, weights
them by their clicks and draws 50,000 clicks with replacement (39,150 distinct queries), of which 40.4\% are navigational. \trafficjul{}
applies the same construction to July 2026, inside the training period. It is our development set: every threshold and weight in this
section was chosen on it.

A click table already answers much of this traffic, so we divide navigational traffic by what such a table can know
(\cref{fig:blocks}). \emph{Seen pairs}, whose query and target occur together in the training data, carry 62.0\% of it. \emph{New
queries}, absent from the training data, carry 33.1\%. We divide them by their target: \emph{familiar} targets have at least 11 training
pairs, \emph{cold} targets have 1 to 10, and targets outside \Ufull{} are missed by every system. The threshold of 11 is where GR
overtakes dense retrieval on new queries, both here and on the held-out queries below (\cref{app:newq,app:heldnav}). The remaining 4.9\%
are \emph{other} traffic: queries that occur in the training data with a different target, and queries that we withheld from SID training to build \heldnav{} (below) but that the click table still contains. In all, the target lies in \Ufull{} for 79.2\% of navigational traffic.

New queries in traffic mix two effects, since most of their targets are outside the universe or rarely clicked. \heldnav{} isolates
one of them, generalisation to a new query for a URL that is in the universe (\cref{sec:heldnav}). Its 60,000 navigational queries of
the training period were drawn in equal numbers from three bands of query clicks and removed from SID training with their pairs, and
the click table has no entry for them. We exclude queries whose target is outside \Ufull{} and those that occur in the July data behind GR's URL-string pre-training and the dense retriever's maps. This leaves 12,967 queries, each weighted once (\cref{app:heldnav}).

We report navigational hit@1 and hit@10 (\cref{eq:hitk}) on \trafficsep{}, weighted by clicks; \cref{app:macro} weights every query once
instead. Differences are in percentage points and, like other derived quantities, are computed from unrounded values, so they can differ by 0.1 from what the rounded numbers suggest. Their 95\% confidence intervals, clustered by query, are about $\pm2.4$ points for a single
hit@1 on \trafficsep{} and $\pm0.2$ to $\pm2.8$ for paired differences between systems, and \cref{tab:ci} gives the interval of every
difference quoted in the text. In the tables, bold marks the best value among the systems or training runs being compared, by column or,
where they form the columns, by row. Reference rows (marked $^\dagger$), ceilings and untrained starting points are not compared.

\subsection{Baselines}
\label{sec:baselines}

\emph{Dense retrieval} keeps Octen-Embedding-4B~\citep{octen2025embedding} frozen and embeds queries and URLs with it. The first 1,024
dimensions pass through one linear map per side, $W = I + \Delta W$, before an exact search of every URL by cosine similarity. Only the
maps, 2.1M parameters, are trained, contrastively on navigational query--URL pairs from our click data, and the final maps
average seven runs (\cref{app:densecfg}). The URL-side map also gives the embeddings from which the SIDs are built (\cref{sec:sid}).
A similarity score knows nothing of how popular a page is, which matters for navigation, so \emph{dense + click prior} reranks the top
100 by cosine similarity plus $0.04\log_{10}(1 + n_u)$, where $n_u$ is the number of training clicks on URL $u$. \emph{BM25} (Lucene,
$k_1=0.9$, $b=0.4$, untuned) indexes the same URL texts that the dense retriever encodes.

The \emph{click-table lookup} ranks the URLs on which the query was clicked up to August 2026 by their clicks. It is a reference for seen
queries rather than a competing system. \emph{Lookup$\rightarrow$GR} and \emph{Lookup$\rightarrow$dense} return, for every query in the
click table, the lookup's ranking followed by the other system's results, and that system's results otherwise. Six frontier LLMs answer
a 5,000-query sample of \trafficsep{} from memory (\cref{sec:llmres}), and six public encoders are compared on 32M-URL subsets
(\cref{sec:heldnav,app:dense}).

\subsection{Main results}
\label{sec:mainres}

\begin{table}[t]
  \centering
  \small
  \setlength{\tabcolsep}{3.5pt}
  \caption{\textbf{Navigational hit@1 / hit@10 (\%) on \trafficsep{}}, overall and by block, with each block's share of navigational
  traffic: seen pairs; new queries whose target in \Ufull{} has at least 11 training pairs (familiar) or 1 to 10 (cold); and other
  traffic, queries that occur in the training data with another target (changed) or that were withheld from SID training but are in
  the click table (withheld; by clicks, 88.5\% of them, mostly head queries, also occur in the July data behind GR's URL-string pre-training). New queries whose target is
  outside \Ufull{} (18.6\%) are a miss for every system. Every system returns URLs from \Ufull, except $^\ddagger$\modelsmall, which returns URLs from \Usmall{} (it
  contains the target for 68.9\% of navigational traffic). $^\dagger$Uses the training click table directly: a reference for seen
  queries, not a competing system. Dense retrieval (exact) scores the query against every URL vector, without an approximate index, and
  dense + click prior reranks its top 100 by the URLs' training clicks (\cref{sec:baselines}). $^\S$Fills the rest of the lookup's list,
  and answers the queries that the table does not contain, with GR or dense retrieval as chosen by the two-threshold router of \cref{sec:combo}.
  Bold: best single system and best lookup-first combination per column.}
  \label{tab:main}
  \begin{tabular}{lcccccc}
\toprule
 & & & \multicolumn{2}{c}{New query, target in $\mathcal{U}_{\text{201M}}$} & \multicolumn{2}{c}{Other} \\
\cmidrule(lr){4-5}\cmidrule(lr){6-7}
 & All & Seen pair & Familiar & Cold & Changed & Withheld \\
 & (100.0\%) & (62.0\%) & (9.0\%) & (5.5\%) & (3.9\%) & (1.0\%) \\
\midrule
Octen-WebGR-4B & \textbf{59.6} / \textbf{68.7} & \textbf{88.2} / \textbf{96.3} & \textbf{30.9} / \textbf{58.2} & 14.3 / 32.3 & 9.5 / 24.1 & \textbf{92.0} / \textbf{100.0} \\
Octen-WebGR-0.6B$^\ddagger$ & 56.5 / 64.0 & 84.4 / 91.6 & 30.2 / 55.5 & 3.4 / 7.9 & 8.9 / 21.7 & 91.0 / 99.0 \\
Dense retrieval (exact) & 24.8 / 55.6 & 34.4 / 75.9 & 17.2 / 45.9 & 24.3 / \textbf{52.3} & 6.8 / 20.5 & 33.0 / 73.5 \\
Dense + click prior & 46.5 / 64.2 & 68.7 / 89.2 & 21.4 / 53.9 & 13.1 / 38.3 & \textbf{11.3} / \textbf{24.5} & 83.5 / 94.0 \\
BM25 (URL + queries) & 9.1 / 25.3 & 10.8 / 31.3 & 9.2 / 27.4 & \textbf{25.5} / 49.8 & 1.9 / 12.2 & 5.0 / 26.5 \\
Click-table lookup$^\dagger$ & 58.9 / 62.5 & 93.4 / 99.2 & 0.0 / 0.0 & 0.0 / 0.0 & 0.0 / 0.0 & 99.5 / 100.0 \\
\midrule
Lookup $\rightarrow$ GR & 62.5 / 70.2 & 93.4 / 99.2 & 30.9 / 58.2 & 14.3 / 32.3 & 0.3 / 18.9 & 99.5 / 100.0 \\
Lookup $\rightarrow$ dense & 61.8 / 70.1 & 93.4 / 99.2 & 17.2 / 45.9 & \textbf{24.3} / \textbf{52.3} & \textbf{0.6} / 16.2 & 99.5 / 100.0 \\
Lookup $\rightarrow$ routed$^\S$ & \textbf{63.1} / \textbf{71.0} & 93.4 / 99.2 & \textbf{32.8} / \textbf{58.3} & 22.4 / 45.0 & 0.5 / \textbf{19.3} & 99.5 / 100.0 \\
\midrule
Target in $\mathcal{U}_{\text{201M}}$ (ceiling) & 79.2 & 99.4 & 100.0 & 100.0 & 51.6 & 100.0 \\
\bottomrule
\end{tabular}

\end{table}

GR is the most accurate system on navigational traffic.
\model{} ranks the target first for 59.6\%, against 24.8\% for dense retrieval and 9.1\% for BM25 over the same
URL texts (\cref{tab:main}). At hit@10 the gap to dense retrieval narrows, 68.7\% against 55.6\%. The difference between dense
retrieval's hit@10 and hit@1 is the 30.8\% of navigational traffic for which it has the target in its top ten but ranks another URL
first. For 46.1\% of this traffic, that URL is another page of the target's site, such as a help page, a login page or a version in
another language, and for the rest an encyclopedia article, a similarly named domain or another site (\cref{tab:denseerr}). GR ranks
the target first for 81.5\% of this traffic.

A popularity prior closes most of the gap at hit@1. Reranked by the URLs' training clicks, dense retrieval reaches 46.5\% hit@1, and GR
leads it by 13.0 points. What dense retrieval lacks is a score term that depends on the URL alone; GR learns this term from the clicks,
and on \trafficjul{} adding the same prior to GR does not help. GR's lead is largest on queries with at least 1,000 September clicks,
56\% of navigational traffic, where it ranks the target first for 90.4\%, against 72.7\% for dense retrieval with the click prior
and 33.1\% without.

Serving GR costs no more than serving dense retrieval. Per query, \model{} needs 2.3 GPU-milliseconds with our decoding code, which is
not part of the release (the released reference implementation needs 10.1 in batches of 32), and dense retrieval an estimated 3.9.
GR's index is its weights and identifier keys, about 10\,GB of GPU memory, whereas exact dense search keeps 412\,GB of vectors in
memory (\cref{tab:models,sec:serving}). GR's cost lies in training instead: the SID training of \model{} took about 1,000 H200
GPU-hours, against about 23 for the dense retriever's maps.

GR's gain over a click table lies in new queries for familiar pages. On seen pairs, \model{} is correct at rank one for 88.2\% of
the traffic, against 93.4\% for a lookup in the training click table. The lookup is the procedure closest to how targets are defined, and
on these pairs GR behaves as a lossy copy of the table. The table cannot answer new queries at all. In the familiar
block, new ways of asking for pages with many training queries, GR ranks the target first for 30.9\%, against 17.2\% for dense
retrieval and 21.4\% with the click prior. On cold targets the order reverses. Dense retrieval and BM25 rank the target first for 24.3\%
and 25.5\%, GR for 14.3\%, and the click prior, which favours popular URLs, lowers dense retrieval to 13.1\%.

Behind a click table, GR adds a little more than dense retrieval does. Over all traffic GR alone is slightly ahead of the lookup
(59.6\% against 58.9\%). Serving the lookup first yields 62.5\% with GR filling the rest and 61.8\% with dense retrieval. Of this
difference of 0.7 points of all navigational traffic, the familiar block contributes $+1.2$ and the cold block takes 0.6 away. In the
other blocks the lookup decides. Where a query's target has changed since August, the lookup serves the old target first (0.3\% hit@1,
against 9.5\% for GR alone). Withheld queries it answers almost always, and so does GR (92.0\% of their traffic). \Cref{sec:newpages} traces GR's weakness on cold targets to how familiar the model is
with the page.

\subsection{Held-out queries}
\label{sec:heldnav}

\begin{table}[t]
  \centering
  \small
  \setlength{\tabcolsep}{3pt}
  \caption{\textbf{\heldnav: hit@1 / hit@10 (\%) on 12,967 navigational queries withheld from training}, every query weighted once:
  overall, by the target's training pairs (familiar: at least 11) and by the query's clicks (head: at least 1,000; torso: 100 to 999;
  tail: fewer than 100), with numbers of queries in parentheses. The click table answers none of them. URL texts are cleaned of the
  withheld queries (\cref{app:heldnav}). $^\ddagger$Returns URLs from \Usmall. $^\S$The router of \cref{tab:main}. Bold: best single
  system per column.}
  \label{tab:heldnav}
  \begin{tabular}{lcccccc}
\toprule
 & All & Familiar & Cold & Head & Torso & Tail \\
 & (12,967) & (6,513) & (6,454) & (1,881) & (5,981) & (5,105) \\
\midrule
Octen-WebGR-4B & \textbf{32.0} / \textbf{54.2} & \textbf{49.5} / \textbf{74.2} & 14.3 / 34.1 & \textbf{69.8} / \textbf{85.4} & \textbf{26.8} / \textbf{48.9} & \textbf{24.2} / \textbf{49.0} \\
Octen-WebGR-0.6B$^\ddagger$ & 27.5 / 42.7 & 48.7 / 72.8 & 6.1 / 12.4 & 66.9 / 85.2 & 22.1 / 36.6 & 19.3 / 34.2 \\
Dense retrieval (exact) & 20.6 / 49.5 & 19.5 / 53.3 & 21.8 / 45.7 & 25.6 / 63.3 & 21.5 / 48.5 & 17.8 / 45.6 \\
Dense + click prior & 20.8 / 47.2 & 31.5 / 64.2 & 10.1 / 30.1 & 48.4 / 77.4 & 17.8 / 43.6 & 14.2 / 40.4 \\
BM25 (URL + queries) & 16.5 / 39.4 & 7.8 / 31.2 & \textbf{25.3} / \textbf{47.6} & 9.0 / 33.4 & 18.4 / 40.5 & 17.0 / 40.3 \\
\midrule
Routed$^\S$ & 35.1 / 58.2 & 49.8 / 73.5 & 20.3 / 42.7 & 71.7 / 87.7 & 30.8 / 54.2 & 26.7 / 51.9 \\
GR $\cup$ dense, top 10 & -- / 68.6 & -- / 80.1 & -- / 57.0 & -- / 92.4 & -- / 65.1 & -- / 64.0 \\
\bottomrule
\end{tabular}

\end{table}

On queries withheld from training, GR again wins on familiar pages and loses on cold ones. A withheld query can still reach a
system through four other channels, which we close or measure (\cref{app:heldnav}). The URL texts that BM25 indexes and the dense
retriever encodes contain the query for 77\% of the test queries; they were rebuilt without it, which lowers the hit@1 of dense retrieval
from 35.7\% to 20.6\%. The July data behind GR's URL-string pre-training and the dense retriever's maps contain 47\% of the queries, which are excluded.
GR's identifiers and indexing samples were built from the uncleaned texts, and two controls show no effect on GR. On the remaining
queries, \model{} ranks the target first for 32.0\%, against 20.6\% for dense retrieval, 20.8\% with the click prior and 16.5\% for BM25
(\cref{tab:heldnav}). On familiar targets GR reaches 49.5\%, against 19.5\% for dense retrieval. On cold targets it reaches 14.3\%,
against 21.8\% for dense retrieval and 25.3\% for BM25. \modelsmall{} matches the 4B model on familiar targets (48.7\%) but finds few cold
ones, most of which lie outside its smaller universe.

Off-the-shelf embedding models show the same split from the other side. We evaluate them on a 32.4M-URL subset: the 32M subset
of \cref{sec:ladder}, extended with every \heldnav{} target and the 50 most-clicked pages of its host (\cref{tab:heldnavzs}). Used
zero-shot, the six encoders from 0.6B to 8B parameters rank the target first for 29.2--32.4\% of the queries, against 39.8\% for GR and
34.8\% for the dense baseline. On familiar targets they reach only 19--21\%, against 57.3\% for GR. On cold targets they are the best
systems we measured, 39--45\% against 35.2\% for the dense baseline and 22.1\% for GR, and training linear maps on navigational pairs
moves them towards popular pages. GR thus generalises on the query side, to new ways of asking for pages it has learned, and
general-purpose embeddings on the page side, to pages that few queries have reached.

\subsection{Combining GR with dense retrieval}
\label{sec:combo}

The two systems err on different queries, and two signals
available at serving time tell many of them apart: the number of training pairs of GR's first URL, and GR's own beam probability. We
route a query to GR when its first URL has at least three training pairs and a probability of at least 0.15, and to dense retrieval
otherwise, with both thresholds chosen on \trafficjul. Behind the lookup, this adds 0.62 points of hit@1 (63.1\% against 62.5\%); either
test alone adds 0.4. The router keeps GR's lead on familiar targets (32.8\%) and recovers most of dense retrieval's on cold ones (22.4\%
against 24.3\%). Interleaving the two lists, GR first, adds 1.6 points of hit@10, and the union of both top-10 lists contains the target
for 72.6\% of navigational traffic, against a ceiling of 79.2\%. On \heldnav{} the same router reaches 35.1\% hit@1, against 32.0\% for
GR alone, and the union of the two top-10 lists 68.6\% (\cref{tab:heldnav}). A general-purpose encoder is the better partner. On the
32.4M-URL subset, the union of GR's top ten with that of Qwen3-Embedding-8B, used zero-shot, contains the target of 82.1\% of \heldnav{}
queries, against 78.9\% with the dense baseline and 62.5\% for GR alone (\cref{tab:heldnavzs}).

\subsection{Closed-book LLMs}
\label{sec:llmres}

Frontier LLMs answering from memory fall far short of GR. The strongest frontier models, GPT-6 Astra and Claude Opus 5.5, rank
the target first for 37.5\% and 36.8\% of the sample, against 59.8\% for \model{} on the same queries (\cref{fig:cost,tab:llm}). They
often identify the site (52.4\% and 52.7\% at the host level), which is still below GR's 59.8\% for the exact URL, and on the sample's
new queries they are below GR as well (8.4\% hit@1 for Astra against 11.5\%; \cref{app:llm}). Relative to dense retrieval they are
better at hit@1 and worse at hit@10: they commit to one plausible site but rarely list the right alternatives.

\subsection{A public benchmark}
\label{sec:public}

ORCAS-Nav, 20,000 navigational queries from public Bing clicks, weights rarely clicked targets far more than traffic does
(\cref{app:public}). GR leads dense retrieval at hit@1 (16.2\% against 11.8\%) but trails it with the click prior (21.0\%), in line with
dense retrieval's advantage on cold targets.

\section{Scaling the Number of URLs and the Model}
\label{sec:scaling}

How does GR behave as it grows? We vary two things, the number of URLs a model may return and the number of its parameters.
\Cref{sec:ladder} varies the universe at decoding time with the model fixed, \cref{sec:s1} varies the model on fixed data,
\cref{sec:grid} varies both at a fixed number of exposures per query--URL pair, and \cref{sec:smalluniverse} matches the universe to
the model. Only \cref{sec:ladder} compares GR with dense retrieval, because how much accuracy a system loses as URLs are added means
little without a reference. The other experiments compare GR models with one another.

\subsection{Number of candidate URLs}
\label{sec:ladder}

\begin{figure}[t]
  \centering
  \includegraphics[width=\linewidth]{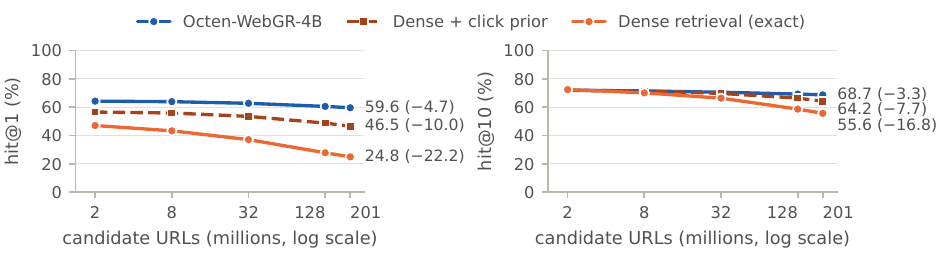}
  \caption{\textbf{Accuracy against the number of candidate URLs.} The same GR model and the same dense index, with and without the
  click prior, restricted to random subsets of \Ufull{} that contain every target and its same-host neighbours. Labels give the value at 201M URLs and, in
  parentheses, the change from 2M.}
  \label{fig:dladder}
\end{figure}

Adding candidate URLs costs GR far less accuracy than dense retrieval. We restrict the released model and the dense index to
nested random subsets of \Ufull{} with 2M, 8M, 32M and 128M URLs. All subsets contain the navigational targets of \trafficsep{} and their
same-host neighbours (\cref{app:ladder}). GR decodes over the trie of the subset, and dense retrieval searches only the subset. From 2M
to 201M URLs, GR hit@1 decreases by 4.7 points, from 64.2\% to 59.6\%, and dense hit@1 by 22.2 points, from 47.0\% to 24.8\%
(\cref{fig:dladder}). Reranked with the click prior, dense retrieval loses 10.0 points, from 56.5\% to 46.5\%, still 5.3 points more than
GR. The two systems differ in what an added URL does. For dense retrieval it is one more URL that can come close to the query by
chance (\cref{sec:background}). GR has concentrated its probability on the URLs users click, so most added URLs never reach
the top ranks, as an earlier analysis predicted~\citep{zhang2025grvsdr}.

\subsection{Model size at fixed data}
\label{sec:s1}

Larger models are ahead at every point of training, but in a fixed number of steps 8B gains little over 4B. The 0.6B, 1.7B, 4B
and 8B Qwen3 base models were trained from their base checkpoints on the same sample sequence over \Ufull{} for 6,000 steps (1.6B tokens)
with the same schedule (\cref{tab:s1sep}). At step 6,000, 8B leads 4B by only 0.77 points of hit@1 on \trafficsep{}, and on new queries all four stay between 4.7\% and 6.6\%. Runs of a
fixed number of steps, however, tie model size to how often each pair is seen. The next experiment fixes the number of exposures
instead.

\subsection{Model size at a fixed number of exposures}
\label{sec:grid}

Larger models learn more from each exposure. Training on more URLs
entails more samples, longer runs and more forgetting, so we compare five Qwen3 base models, from 0.6B to 14B parameters, at a fixed number of exposures per pair. We draw nested
random subsets of 2M, 8M and 32M URLs from \Ufull{} and keep their SIDs. Each URL is represented by its two most distinctive query pairs,
its primary and secondary pair, and each pair is trained exactly twice, so that the runs see 8M, 32M and 128M samples in 441, 1,764 and 7,058 steps. Each model first learns the task on a disjoint 8M-URL subset, which gives its starting point, and is then trained at a constant learning rate of $3.5\times10^{-5}$ with a decay over the final 10\%
of steps, and evaluated on the same 20,000 trained pairs (\cref{app:grid}). Besides hit@1, we measure what training taught: how much it
lowered the negative log-likelihood (NLL) of the gold SID from its value at the starting point, which we call the zero-shot NLL, in nats per pair. A reduction of one nat makes the gold
identifier $e \approx 2.7$ times as likely as at the starting point.

\begin{figure}[t]
  \centering
  \includegraphics[width=\linewidth]{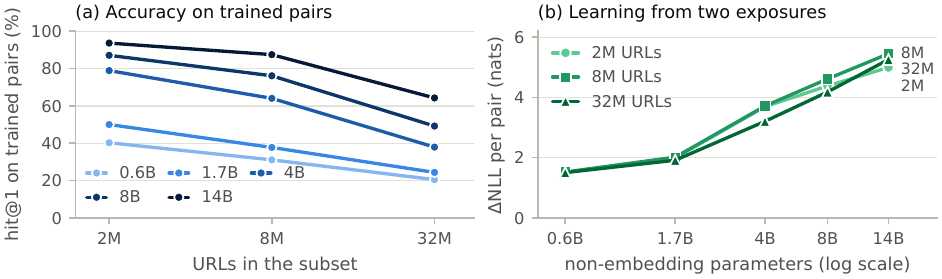}
  \caption{\textbf{The fixed-exposure grid}, in which each URL's two query pairs are each trained twice. (a) hit@1 on the trained pairs,
  one line per model size. (b) Reduction in the NLL of the gold SID per pair against non-embedding parameters on a logarithmic axis, one
  line per subset: larger models learn more from each exposure at every number of URLs, and the three lines nearly coincide, which is the
  result: how much an exposure teaches depends on the size of the model, not on the number of URLs.}
  \label{fig:grid}
\end{figure}

Accuracy decreases as URLs are added (\cref{fig:grid}a): the 4B model ranks 78.9\% of the trained pairs first on 2M URLs and 37.9\% on
32M. Most of the decrease reflects a harder task rather than forgetting. From 2M to 32M URLs the trained NLL of the gold SID rises by
1.7--2.4 nats for every model. Of this, 1.9--2.0 nats is the rise of the zero-shot NLL at the second identifier level, where more URLs
share each first-level code; the gain from training changes by at most half a nat (\cref{sec:forgetting}). The NLL reduction per pair
increases with model size at every number of URLs (\cref{fig:grid}b). On 8M URLs it is 1.5 nats at 0.6B, 2.0 at 1.7B, 3.7 at 4B, 4.6
at 8B and 5.5 at 14B, and from 4B upward each doubling of parameters adds 0.7--1.3 nats. Larger models therefore index more URLs at the
same accuracy: the 4B model on 32M URLs is as accurate as the 1.7B model on 8M (37.9\% and 37.7\% hit@1). Fits over the grid equate
four times the URLs with about 2.5 times the parameters (\cref{app:grid}). This rule mostly measures how much more a model must learn
per exposure to separate URLs in the denser first-level groups of our two-token identifiers.

\subsection{Matching the universe to the model}
\label{sec:smalluniverse}

A small model gains from a smaller universe, but only when it is both trained and decoded there. At a fixed number of steps, a
smaller universe spends its samples on fewer pairs, and more of them are the pairs that carry the most traffic. This benefits a model
that learns little from each exposure. Repeating the 6,000-step runs on \Usmall{} raises hit@1 by 10.2, 8.9 and 7.6 points at 0.6B, 1.7B
and 4B (\cref{app:smalluniverse}), although \Usmall{} contains the target for only 68.9\% of the traffic, against 79.2\% for \Ufull. The
advantage shrinks with training and, up to 4B, with model size. Decoding over a smaller universe alone does not help. Restricting the
models trained on \Ufull{} to \Usmall{} changes them by at most half a point, and decoded over \Ufull{} the models trained on \Usmall{}
keep only 3.9, 4.9 and 3.1 of their 10.2, 8.9 and 7.6 points. Part of the gain comes through the
sample mix. The navigational floor covers every navigational query of a universe, so a larger universe gives it a
larger share of the samples, and a smaller one leaves a larger share to the click-weighted samples that carry the traffic.

\modelsmall, trained on \Usmall{} for 43.9K steps (\cref{sec:smalltrain}), reaches 56.5\% hit@1. That is 3.1 points below \model{} with a
sixth of the parameters, and level with it where its universe contains the target (81.9\% against 82.3\%). On \Ufull, the
same recipe gives 51.5\% (\cref{tab:smalluniverse}). The best universe thus grows with the length of training and, where a larger model learns more,
with its size.

\section{What Limits a Model: Exposure versus Capacity}
\label{sec:limits}

What limits a model of a given size? It may be limited by \emph{capacity}, if its parameters cannot hold more query--URL pairs, or by
\emph{exposure}, if pairs are not seen often enough, or late enough, to survive training. The answer decides how to scale GR: a capacity
limit calls for larger models, an exposure limit for different training. This section answers the question with further experiments,
on the grid of \cref{sec:grid} and on the training rounds of \model. Within the range we measure, the evidence points to exposure
(\crefrange{sec:forgetting}{sec:bits}). \Crefrange{sec:expo}{sec:newpages} then show where exposure can stand in for parameters and
where it cannot.

\subsection{Forgetting rather than crowding out}
\label{sec:forgetting}

\begin{figure}[t]
  \centering
  \includegraphics[width=\linewidth]{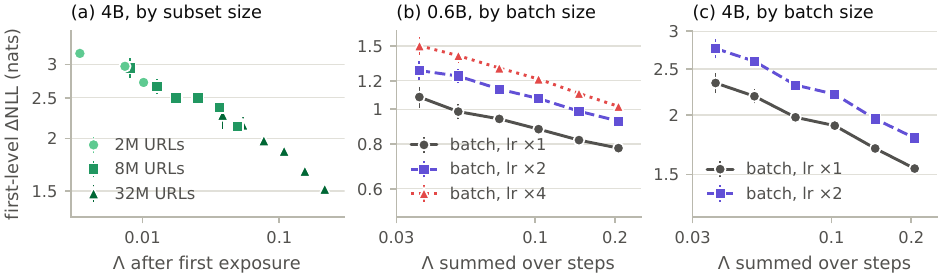}
  \caption{\textbf{Forgetting, on logarithmic axes.} (a) First-level gain per pair (zero-shot minus trained NLL of the first SID token)
  of the 4B grid model against the cumulative learning rate $\Lambda$ after the pair's first exposure, for pairs whose two exposures fell
  in the constant-rate phase; binned, $\pm$ one standard error, subsets offset slightly sideways. (b, c) The same gain on the 32M-URL
  subset trained with a larger batch and the learning rate scaled in proportion, against $\Lambda$ summed over steps. The lines are
  parallel: a larger batch starts higher, because each exposure teaches more, but decays at the same rate.}
  \label{fig:mech}
\end{figure}

What a pair retains depends on the training that follows it, not on how many other pairs the model holds. If parameters were
the limiting factor, every added URL would leave less room for each pair. We test this at the first identifier level of the grid (\cref{sec:grid}), because it poses the same question on every subset: the first code is always one of the same 65,536, whereas the second must tell apart the URLs of a group, which grow more numerous with the subset. The gain at the first level decreases by about a fifth to a third from 2M to 32M URLs (the second
level offsets the loss; \cref{app:forgetting}). Samples are presented in a fixed random order, so where a pair's exposures fall is independent of the pair, and $\Lambda$
(\cref{eq:cumlr}) measures how much training followed an exposure. Plotted against $\Lambda$ after a pair's first exposure, the
first-level gain falls close to a single curve whatever the number of URLs (\cref{fig:mech}a). The decisive comparison fixes both
exposures. We restrict to pairs whose two exposures fell in a common part of the constant-rate phase, so that the training after them
has the same distribution. Going from 8M to 32M URLs then changes the first-level gain by $+0.12$ nats pooled over five models, against
raw changes of $-0.09$ to $-0.67$: once the training that follows a pair is equal, more URLs cost it nothing.

\subsection{The decay window}
\label{sec:window}

A pair is retained best when it is exposed late in training, at a learning rate still high enough to learn it. In full training
runs the same erosion decides what the end point retains. Plotted against $\Lambda$ after a pair's last exposure, a pair stays in the
top ten up to $\Lambda\approx0.29$, is half forgotten by $\Lambda\approx0.42$ and is forgotten above about 0.56 (measured in round 1).
Steps are the wrong unit: pairs early in the window of round 3, followed by 5,900 steps but little learning rate, are retained as well as
later ones (\cref{app:window}). Since weight decay is zero, this forgetting is interference from later updates rather than the
exponential averaging of updates that weight decay induces~\citep{wang2025weightdecay,bergsma2025straight}.

The final decay window is where little learning rate follows an exposure. The windows of rounds 3 and 4 are curated (\cref{sec:rounds}),
and part of the 4.0-point gain from round 2 to round 3 is due to that curriculum. Their effect on retention therefore has to be measured
on pairs that the curation left alone, which land inside or before the window by the random order of samples. We sample 20,000 training
pairs whose query sends at least half of its clicks to one URL, so that each pair has a single answer. Among those that the curation
left alone, the end point of round 3 ranks first 45.3\% of the pairs drawn inside the window and 30.0\% of those drawn before it, a
difference of 15.3 points; when the pairs are drawn by clicks instead, the difference shrinks to 1.8 and is within noise
(\cref{app:window}). In round 4, whose exponential decay spans 74\% of the round, the same comparison shows no difference.

What an exposure teaches also depends on the learning rate at which it is made. The two arms of round 4 (\cref{sec:schedule}) train the
same samples in the same order. Consider old-page pairs drawn once in the first fifth of the window. The learning rate that follows is
the same in both arms ($\Lambda$ of 0.45 and 0.44), but the released arm makes the exposure at 0.79 times the peak and the constant-rate
arm at 0.35 times. Over the end point of round 3, the first gains 6.2 points of hit@1 on \trafficjul{} and the second 0.3. Exposures at
0.2 times the peak or less, at the end of the released arm's window, gain almost nothing.

\subsection{Batch size}
\label{sec:batch}

Forgetting follows the learning rate summed over steps, not over samples. This is expected if forgetting follows how far the parameters move, since Adam's update per step is roughly proportional to the
learning rate whatever the batch size. We retrained the 0.6B grid model on its 32M-URL samples with half to four times the batch size,
and the 4B model with twice the batch size. As a function of $\Lambda$ summed over steps, the share of the first-level gain that
survives between $\Lambda=0.035$ and $0.21$ is the same in every arm (0.67--0.72 at 0.6B, 0.65--0.66 at 4B; \cref{fig:mech}b,c). A window
of a given length in steps therefore holds as many exposures as the batch allows. With a proportionally scaled learning rate each
exposure also teaches more: twice the batch size raises the 4B model's hit@1 in the grid from 37.9\% to 43.5\%, and four times the batch
size the 0.6B model's from 20.5\% to 22.9\% (\cref{app:batch}).

\subsection{Storage efficiency}
\label{sec:bits}

How much does \model{} store? We credit each of the 20,000 single-answer pairs of
\cref{sec:window} (about 420 million of the 1.13 billion training pairs qualify) with the NLL reduction of its gold SID from the
starting point of the grid, which knows the task but not these pairs, to \model: 6.9 bits per pair, or 0.66 bits per parameter
(\cref{app:bits}). Published estimates are 2 bits per parameter for synthetic facts seen 1,000 times each, about 1 when each is seen
100 times~\citep{allenzhu2025capacity}, and 3.6 for random sequences trained to saturation~\citep{morris2026memorize}. Most of our pairs
are seen only a few times, and more exposures raise the bits stored per parameter (\cref{sec:expo}). This places \model{} far from known
ceilings without ruling capacity out.

\begin{figure}[t]
  \centering
  \includegraphics[width=\linewidth]{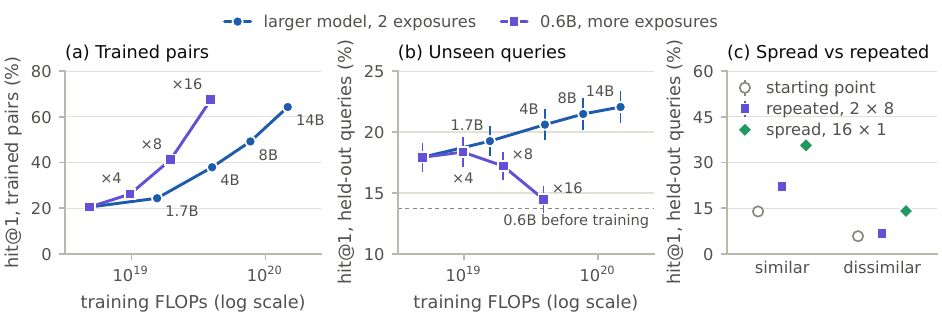}
  \caption{\textbf{More exposures against more parameters} on the 32M-URL subset of the grid. Violet: the 0.6B model trained for 2, 4, 8
  and 16 exposures per pair; blue: models from 0.6B to 14B trained twice per pair; the two series share their first point, and FLOPs
  count the training on the subset only. (a) hit@1 on the trained pairs. (b) hit@1 on 4,000 navigational queries withheld from training
  whose targets are in the subset, with 95\% confidence intervals; the dotted line is the 0.6B model before training on the subset. (c)
  The 3.8M URLs of the subset with at least 16 training pairs, given 16 samples each: the grid's two pairs eight times each (repeated) or
  16 different pairs once each (spread). hit@1 on the 2,597 held-out queries for these URLs, split by whether a query is similar to one of
  its URL's 16 training queries (\cref{app:spread}), with 95\% confidence intervals.}
  \label{fig:expo}
\end{figure}

\subsection{More exposures per pair}
\label{sec:expo}

On trained pairs, repeating exposures can substitute for parameters. The grid trains every pair twice, and at two exposures the
0.6B model stores only 0.19 bits per parameter on 32M URLs, too little to come near a capacity limit. We therefore trained it from the
same starting point on the same samples, replayed in the same order, for 4, 8 and 16 exposures per pair (\cref{app:expo}). The gain per primary pair rises from 1.50 to 2.27, 3.63 and 5.39 nats. The bits stored per parameter grow by about 1.5 times per doubling, from 0.19 to
0.69: no limit lies below 0.69 bits per parameter, although the ladder does not show where one lies. After 16 exposures the 0.6B model
ranks 67.4\% of its trained pairs first, more than the 14B model trained twice per pair (64.3\%), with about a quarter of the FLOPs and in
the same 14.4 hours on eight H200 GPUs (\cref{fig:expo}a).

\subsection{Repeated versus spread exposures}
\label{sec:spread}

For queries the model has not seen, spreading exposures over a page's queries helps, and repeating the same queries does not. On
4,000 navigational queries withheld from training whose targets are in the subset, hit@1 falls from 17.9\% at two exposures to 14.5\% at
16, barely above the 13.7\% of the starting point, which has trained on no pair of the subset. With two exposures, by contrast, it rises
with model size, to 22.1\% at 14B (\cref{fig:expo}b). The cost lies in repeating the same queries. On the 3.8M URLs of the subset with at
least 16 training pairs, we gave every URL 16 samples in the same positions of a run: either its two pairs eight times each, or 16
different pairs once each. Spread over 16 queries, the samples raise hit@1 on the 2,597 held-out queries for these URLs from 18.6\% to
30.6\% (12.0\% before training; \cref{fig:expo}c). Most of the gain is on held-out queries similar to one of the URL's training queries,
77\% of them ($+13.5$ points); on queries similar to none it is $+7.3$ (from 6.8\% to 14.0\%; \cref{app:spread}). Repeating a URL's two
pairs fits those two queries. Spreading the same samples over its queries reaches close variants of them and, less often, queries unlike
any of them.

\subsection{Retrieval of unfamiliar pages}
\label{sec:newpages}

GR's weakness on new pages comes from how little it has seen of them, not from worse identifiers. \model{} ranks the target
first for 26.9\% of the traffic to new pages and for 78.0\% of the traffic to old ones. A new page has a median of 2 training pairs,
against 98 for an old page. In a 4B model trained for 6,000 steps from its base checkpoint, a pair drawn once is ranked first 44.2\% of
the time for an old page and 10.1\% for a new one, but pages drawn equally often as a training target, 0, 1 or 2--3 times, differ by only
2--5 points (\cref{app:newpages}). Familiarity comes from a page's queries. Document-indexing samples (\cref{sec:samples}) raised hit@1 for
new pages by 8.5 points in a randomised comparison, whereas for the 4B model a sample that carries only the URL teaches a tenth of what
two query pairs teach (0.35 against 3.93 nats per pair; \cref{app:newpages}). A new page thus has few queries to spread its exposures
over (\cref{sec:spread}). Most targets outside \Ufull{} are pages of this kind: 19.0 points of navigational traffic go to pages without
earlier clicks, 93\% of them on hosts already in \Ufull.

\section{Towards Larger Universes}
\label{sec:further}

What do these findings mean for indexing more of the web? This section turns them into a training recipe and estimates what a universe
of a billion URLs would take.

\subsection{Training recipe}
\label{sec:recipe}

The measurements of \cref{sec:limits} suggest five rules for further training rounds.
\begin{enumerate}[leftmargin=1.5em,itemsep=1pt,topsep=2pt]
  \item Expose key query--URL pairs late, while the learning rate is still at least about a quarter of its peak, since exposures
  at lower rates teach little (\cref{sec:window}).
  \item Use the largest batch the hardware allows, with the learning rate scaled in proportion, so that the decay window holds more
  exposures without growing in steps (\cref{sec:batch}).
  \item Avoid re-warming (\cref{sec:schedule}).
  \item Give new pages their queries, not only an identifier (\cref{sec:newpages}).
  \item Evaluate decayed checkpoints only (\cref{app:window}).
\end{enumerate}

\subsection{Projected limits}
\label{sec:cost}

A model of about 13B parameters could index a billion URLs as well as ours indexes 201M, if training delivers the exposures. Three
quantities bound a universe of a billion URLs. We extrapolate from the round-3 end point, whose decay window we measured, and assume that a
larger universe has as many training pairs per URL as \Ufull. \emph{Parameters.} The grid's iso-accuracy rule (\cref{sec:grid}), which
holds the exposures per pair fixed, asks for about 13B parameters and 15 times the training compute of \model. \emph{Exposures.} A decay
window can be about 12,000 steps long before its own start is half forgotten. At our batch size it holds on the order of 200 million
single exposures, about as many pairs as the round-3 end point recalls, and at four times the batch size enough for a billion URLs at
the same recall rate. \emph{Identifiers.} Two tokens of 65,536 codes provide 4.3 billion identifiers, of which a billion URLs would fill 23\% if spread evenly over the first-level groups. URLs cluster, however: with the cap of 16,384 URLs per first-level group used for \Ufull{} (\cref{sec:extend}), the two tokens hold about 1.07 billion URLs. A billion URLs would fill 93\% of them and in practice need a higher cap, at the cost of more URLs away from their nearest centroids, and ten billion URLs a third token or larger codebooks~\citep{chen2025unisearch,liu2026bonsai}. Each figure is
extrapolated from a single point, and scaling trends often fail to extrapolate~\citep{lourie2025unreliable,caballero2023bnsl}. The
exposure budget is moreover an upper bound, since it is measured at hit@10 and exposures late in a window teach little
(\cref{sec:window}).

\section{Deployment and Release}
\label{sec:release}

This section describes the released models, how they are served and how their tables were screened.

\subsection{Released models}
\label{sec:models}

\begin{table}[t]
  \centering
  \small
  \caption{\textbf{The released models.} Parameters include the identifier embeddings. GPU times are measured on idle H200 GPUs (the
  median for single queries) with the released reference implementation and with our decoding code, which reuses the prompt's
  key--value cache; CPU latency is on one socket of an Intel Xeon 8568Y+ (\cref{sec:serving}). hit@1 / hit@10 are on \trafficsep{} navigational traffic (\cref{sec:results}); a target outside
  the model's universe counts as a miss. $^*$SID training only, without the 34K steps of URL-string pre-training before it (\cref{sec:rounds}).}
  \label{tab:models}
  \begin{tabular}{lcc}
    \toprule
     & \model & \modelsmall \\
    \midrule
    Base model & Qwen3-4B-Base & Qwen3-0.6B-Base \\
    Layers / hidden size & 36 / 2,560 & 28 / 1,024 \\
    Parameters, total (non-embedding) & 4.36B (3.63B) & 0.73B (0.44B) \\
    Identifier tokens & $2\times65{,}536$ & $2\times65{,}536$ \\
    URLs in the universe & 201,231,792 (\Ufull) & 31,282,165 (\Usmall) \\
    URLs in the released table & 195,921,090 & 30,931,888 \\
    Training steps / H200 GPU-hours & 147.6K / 1,019$^*$ & 43.9K / 71 \\
    GPU memory & 9.7\,GiB & 1.6\,GiB \\
    GPU latency, single query (reference implementation) & 56\,ms & 43.5\,ms \\
    GPU time per query, batches of 32 (reference implementation) & 10.1\,ms & 3.9\,ms \\
    GPU time per query, batches of 64 (our decoding code) & 2.3\,ms & 1.6\,ms \\
    CPU latency, single query & -- & 72\,ms \\
    \trafficsep{} hit@1 / hit@10 (\%) & 59.6 / 68.7 & 56.5 / 64.0 \\
    \bottomrule
  \end{tabular}
\end{table}

\Cref{tab:models} summarises the two released models. \model{} is Qwen3-4B-Base trained on \Ufull; \modelsmall{} is Qwen3-0.6B-Base
trained on \Usmall, which shares the identifiers of \Ufull{} (\cref{sec:extend}). Their released identifier tables are screened against
blocklists (\cref{sec:screening}) and leave out 2.6\% of the URLs of \Ufull{} and 1.1\% of those of \Usmall; every evaluation in this
report uses the full universes.

\subsection{Serving}
\label{sec:serving}

\model{} answers a single query in 56\,ms on an idle H200 GPU and \modelsmall{} in 43.5\,ms (\cref{tab:models}). In batches of 64,
our decoding code needs 2.3 and 1.6 GPU-milliseconds per query (\cref{fig:cost}; measurements and assumptions in \cref{app:serving}).
The released reference implementation recomputes the prompt for every beam in the second step and pads each batch to its longest
prompt, so in batches of 32 it needs 2.4 to 4.3 times as much GPU time per query. On CPUs, \modelsmall{} answers a query in 72\,ms on one socket of an Intel Xeon 8568Y+, within
0.1 points of the hit@1 of the GPU implementation; the CPU code reuses the prompt's key--value cache and computes the softmax only over
valid continuations.

Per query, GR and dense retrieval need similar compute. Dense retrieval runs a 4B encoder on the query and then scores 201M vectors,
3.9 GPU-milliseconds in all by our extrapolation from an exact search over 17M vectors. What differs is what must be held in memory.
The exact fp16 index takes 412\,GB, more than two H200 GPUs hold. Compressed approximate indexes fit in
6--13\,GB~\citep{jegou2011pq,douze2024faiss}, but return the exact nearest neighbour for only 88--91\% of queries on 100M web
documents~\citep{chen2021spann,jayaramsubramanya2019diskann,chen2024msmarcoweb}. GR's index is its weights, 8.7\,GB in bf16, and the
identifier table. Frontier LLMs answering from memory cost at least 40 times as much per query at list prices and are less accurate.
A search system that also serves informational queries keeps an embedding index, and GR does not replace it. GR can take over the 201M
navigational URLs from that index, and it answers new queries for familiar pages better than the index does. Where a click table is
available, serving it first and routing the rest between GR and dense retrieval gives the highest accuracy we measured (63.1\% hit@1;
\cref{sec:combo}).

\subsection{Released artifacts}
\label{sec:artifacts}

We release the weights of \model{} and \modelsmall, their screened SID-to-URL tables, and inference code for constrained decoding on
GPUs and CPUs. We do not release the training data or any code beyond inference, such as the code that trains the models or builds the
identifiers.
The models and their tables are at \url{https://huggingface.co/Octen/Octen-WebGR-4B} and
\url{https://huggingface.co/Octen/Octen-WebGR-0.6B}. The weights and tables are released for non-commercial use under CC BY-NC 4.0,
and the inference code under the Apache License 2.0. Both models are derived from Qwen3 base models, which are licensed under the
Apache License 2.0.

\subsection{Screening of the released tables}
\label{sec:screening}

We removed every URL whose host, or a parent domain of it, is on the adult, malware, phishing, cryptojacking or stalkerware lists of the
UT1 blacklists~\citep{prigent2026ut1}, every URL listed by URLhaus~\citep{abusech2026urlhaus}, and every URL whose text contains an e-mail
address or a credential. This removes 5,310,702 URLs (2.6\%) from the 4B model's table, mostly from the adult list, and 350,277 from the
0.6B model's table. Screening removes the target of 1.2\% of \trafficsep{} navigational traffic and would lower the hit@1 of \model{} from
59.6\% to 58.7\%. We did not check the tables against lists of child sexual
abuse material, which child-protection organisations share through partnerships~\citep{thiel2023csam,laion2024relaion}, nor screen for
personal pages; both are left to the takedown process.

\subsection{Privacy and intended use}
\label{sec:ethics}

A model that memorises query--URL pairs from click data raises questions that a model of public text does not. Within days of the
release of the AOL query log~\citep{pass2006picture}, one of its users was identified~\citep{korolova2009releasing}, and a released model
can reveal rare training pairs~\citep{carlini2019secretsharer}. Pairs seen once in training are already recalled at rank one 37\% of
the time by a 4B model (\cref{app:s1}). Our safeguards are therefore structural. Removing a row from the table removes a URL from every
possible output, much as search engines delist individual URLs from their results~\citep{bertram2019rtbf}. A new URL, conversely,
becomes retrievable only after training on it (\cref{sec:newpages}). The click data carry no user identifiers, and the released tables
contain URLs and identifiers only. Requests to remove a URL can be sent to \href{mailto:research@octen.ai}{research@octen.ai}, and removals are published as versioned
diffs of the table.

The models are meant for research on generative retrieval and for navigational search over the released URLs. Their knowledge
reflects clicks up to August 2026 and the biases of the rankings under which they were made~\citep{joachims2007clicks}. They should
not be used to infer anything about individual users.

\section{Related Work}
\label{sec:related}

\paragraph{Generative retrieval.} DSI~\citep{tay2022dsi} and NCI~\citep{wang2022nci} generate document identifiers with
sequence-to-sequence models, and GENRE~\citep{decao2021genre} decodes entity names under a prefix trie. Identifiers range from URLs and
quantised codes~\citep{zhou2022ultron} to corpus substrings~\citep{bevilacqua2022seal}, learned codes~\citep{sun2023genret} and the
semantic identifiers of generative recommendation~\citep{rajput2023tiger,zhou2025onerectr}. \citet{pradeep2023scale} scaled GR to the
8.8M passages of MS MARCO, where effectiveness fell with corpus size and an 11B model was slightly worse than a 3B one. Their
encoder--decoders, synthetic queries and identifiers differ from ours, so we cannot attribute the contrast with our results
(\cref{sec:grid}). RIPOR~\citep{zeng2024ripor} and PAG~\citep{zeng2024pag} later matched or exceeded several dense retrievers at that
size. Analytically, GR can be viewed as a form of dense retrieval~\citep{nguyen2023grdr,wu2024grmvdr}, and LTRGR adds a learning-to-rank
stage to it~\citep{li2024ltrgr}.

\paragraph{Generative systems at scale.} Generative and model-as-index systems with 100M or more targets address products
(100M~\citep{chang2021xmcproduct}, over 300M~\citep{ling2026crid}), advertising keywords (295M~\citep{lian2019baidu}) or videos (about
1B~\citep{chen2025unisearch}). In web search, SE-DSI generates official-site URLs for a target set described only as
``moderate''~\citep{tang2023sedsi}, and AuthGR generates host names, with authority scores pre-computed for 3.75M
hosts~\citep{lee2026authgr}. \family{} is, to our knowledge, the first at this scale whose targets are individual web pages. Billion-scale dense
retrieval relies on approximate search over compressed vectors~\citep{jayaramsubramanya2019diskann,chen2021spann,jegou2011pq}, whereas
our dense baseline searches exactly. Navigational search was evaluated as home-page and named-page finding in the TREC Web
Track~\citep{craswell2002trecweb}. Besides ORCAS~\citep{craswell2020orcas}, MS MARCO Web Search pairs a web corpus with click
labels~\citep{chen2024msmarcoweb}.

\paragraph{Scaling.} Generative recommenders saturate early with model size on small catalogues with fixed
data~\citep{liu2026sidscaling}, whereas industrial systems report continued gains~\citep{zhou2025onerecv2,he2025plum}. Dense retrievers
improve with model size~\citep{fang2024denselaws,ni2022gtr} but degrade as the index grows~\citep{reimers2021curse,weller2026limit}.

\paragraph{Memorisation and forgetting.} Language models store about 2 bits per parameter of synthetic facts seen 1,000 times
each~\citep{allenzhu2025capacity} and 3.6 bits per parameter of random sequences~\citep{morris2026memorize}, and memorise more with
size and repetition~\citep{tirumala2022memorization,lu2024factmem}. Pre-training retains mainly what it saw late and forgets at a rate
set by the learning rate~\citep{pagliardini2025ademamix,bergsma2026trec,lesci2024memorisation}. \citet{chang2024acquire} found forgetting
to follow a power law in steps and to slow as the learning rate decays; both follow if forgetting is counted in cumulative learning
rate. GR forgets too. In DSI++, indexing new documents erodes old ones~\citep{mehta2023dsipp}, and later work adds documents by
constrained optimisation~\citep{kishore2023incdsi} or replay~\citep{chen2023clever}, or studies GR on a changing
corpus~\citep{zhang2025dynamic,kim2024dynamic}. These studies count forgetting in steps or corpus increments. We find that its clock is
the learning rate summed over the steps after an exposure, whatever the batch size (\cref{sec:limits}).

\section{Limitations and Future Work}
\label{sec:limitations}

\begin{itemize}[leftmargin=1.5em,itemsep=2pt,topsep=2pt]
  \item \emph{New queries and pages.} New queries make up a third of navigational traffic, and \model{} answers them correctly at rank
  one 10.8\% of the time. More than half of their targets are outside the universe, and on targets with few training queries GR is
  below dense retrieval and BM25. We have not tested indexing samples built from page titles or content.
  \item \emph{Value next to a click table.} Where the click table is served first, GR adds 0.7 points of hit@1 over a dense fallback
  and a two-threshold router another 0.6 (\cref{sec:combo}). A learned router or a reranker over both lists, which could reach the
  72.6\% of their union, is untested. Serving the table first also loses the queries whose target has changed since the table was built
  (3.9\% of traffic): every lookup-first combination ranks the new target first for under 1\% of them, below either model alone.
  \item \emph{Baselines.} The dense baseline trains only two linear maps on a frozen encoder, on navigational query--URL pairs only and with about
  2\% of GR's training compute, and BM25 is untuned. Public encoders are compared only on 32M-URL subsets. A dense encoder fine-tuned in
  full on the same pairs, or a retriever over page content, might narrow the gaps we report.
  \item \emph{Click labels.} Targets are the most-clicked URLs, so they inherit the rankings under which the clicks were made and
  the biases of clicking~\citep{joachims2007clicks}. We have not run an online experiment.
  \item \emph{Scope of the scaling results.} The fixed-exposure grid stops at 32M URLs and two exposures per pair, and its matched
  comparison equalises only the training after a pair. Only the 0.6B model was given more exposures, up to 16. A capacity limit with
  more URLs, larger models or other model families is therefore not excluded, and the projections of \cref{sec:cost}
  extrapolate from a single measured point. Spread exposures were tested only on URLs with at least 16 training queries, nearly all
  old pages.
  \item \emph{Held-out queries and ablations.} \heldnav{} removes withheld queries from the URL texts and excludes those in the July
  data, but GR's identifiers and its document-indexing samples were built from the uncleaned texts. We see no sign that either helps GR
  (\cref{app:heldnav}) but cannot exclude a small effect without retraining. Its queries come from the training period and resemble
  training queries more than truly new ones do, and excluding the July data leaves few head queries. We have not ablated the URL-string pre-training of \model, so how much of its accuracy comes from it rather than from SID training is unknown.
  \item \emph{Benchmarks and reproducibility.} The only public evaluation, ORCAS-Nav, uses clicks six years older than ours, and we
  have not run the recipe on a public click corpus such as MS MARCO Web
  Search~\citep{chen2024msmarcoweb}. The click data are proprietary, and every result comes from a single run per configuration.
  Repeating the 0.6B grid run on 2M URLs with another sample order changes its hit@1 by 0.1--0.2 points.
\end{itemize}

These limitations mark the directions that the measurements of this report make most promising. The first is reaching pages that no
query reaches, for which indexing samples from page titles or content are the untested candidate. The second is delivering more
exposures per decay window through larger batches. The third is testing for a capacity limit beyond 32M URLs, at more exposures and for
larger models. Beyond about a billion URLs, identifiers need a third token or larger codebooks (\cref{sec:cost}).

\section{Conclusion}
\label{sec:conclusion}

\family{} shows that a 4B-parameter generative retriever, together with a table of identifiers, can serve as the navigational index for
201 million web pages, and that a 0.6B model can do the same for 31 million. Its strength lies on the query side. For pages it has
learned from many queries, GR answers new ways of asking for them far better than dense retrieval, both on traffic from the month after
training and on queries withheld from training. Embeddings remain better on pages that few queries reach. The two are complementary,
and even a two-threshold router between them improves on either. Serving GR needs one GPU, at about the compute of the query encoder
that dense retrieval also runs, and no index beyond its weights and a table. Next to an embedding index, it can take over the
navigational URLs. GR's accuracy depends only weakly on how many URLs it may return, and larger models index more URLs at equal
accuracy. What limits it is exposure rather than capacity: the learning rate accumulated over the steps that follow a query--URL pair erodes what
the pair left behind. Scaling further is therefore as much a question of delivering exposures as of training larger models. Pages that
no query reaches remain the open problem.

\bibliographystyle{octen_report}   
{\small\setlength{\bibsep}{2pt}\setlength{\parfillskip}{0pt plus 1fil}\bibliography{refs}}

\clearpage
\appendix
\section{Evaluation Details}
\label{app:evaldetail}

This appendix gives the details behind \cref{sec:results}: the confidence intervals of the differences quoted there, the results with
every query weighted once, the breakdown that separates familiar from cold targets, the construction of \heldnav, the dense encoders and
the dense baseline, and the serving costs.

\subsection{Confidence intervals}
\label{app:ci}

The text quotes differences without their confidence intervals, which \cref{tab:ci} collects. They are 95\% intervals clustered by query;
a single value is the half-width of a symmetric interval.

\begin{table}[H]
  \centering
  \small
  \setlength{\tabcolsep}{4pt}
  \caption{\textbf{Differences quoted in the text, with 95\% confidence intervals}, in points of \trafficsep{} hit@1 unless a row says
  otherwise.}
  \label{tab:ci}
  \begin{tabular}{>{\raggedright\arraybackslash}p{11.6cm}rr}
    \toprule
    Comparison & Difference & 95\% interval \\
    \midrule
    \multicolumn{3}{l}{\emph{\Cref{sec:schedule}: learning-rate schedule}} \\
    Round 4, released arm minus constant-rate arm & $+0.34$ & $\pm0.22$ \\
    \addlinespace
    \multicolumn{3}{l}{\emph{\Cref{sec:mainres}: main results}} \\
    GR minus dense + click prior, all traffic & $+13.0$ & 11.8 to 14.3 \\
    GR minus dense retrieval, familiar block & $+13.7$ & $\pm2.5$ \\
    GR minus dense + click prior, familiar block & $+9.6$ & $\pm2.4$ \\
    GR minus dense retrieval, cold block & $-10.0$ & $\pm2.8$ \\
    GR minus click-table lookup, all traffic & $+0.7$ & $\pm0.4$ \\
    Lookup$\rightarrow$GR minus lookup$\rightarrow$dense, all traffic & $+0.7$ & $\pm0.3$ \\
    \addlinespace
    \multicolumn{3}{l}{\emph{\Cref{sec:heldnav}: \textup{\heldnav}, every query weighted once}} \\
    GR minus dense retrieval, all queries & $+11.4$ & $\pm1.0$ \\
    GR minus dense retrieval, familiar targets & $+30.1$ & $\pm1.4$ \\
    GR minus dense retrieval, cold targets & $-7.5$ & $\pm1.2$ \\
    \addlinespace
    \multicolumn{3}{l}{\emph{\Cref{sec:combo}: combining GR with dense retrieval}} \\
    Behind the lookup, router minus GR & $+0.62$ & $\pm0.15$ \\
    Behind the lookup, interleaved lists, gain in hit@10 & $+1.6$ & $\pm0.2$ \\
    \addlinespace
    \multicolumn{3}{l}{\emph{\Cref{sec:ladder}: from 2M to 201M candidate URLs}} \\
    Change for GR & $-4.7$ & $\pm0.4$ \\
    Change for dense retrieval & $-22.2$ & $\pm1.5$ \\
    Change for dense + click prior & $-10.0$ & $-10.9$ to $-9.1$ \\
    Loss of dense + click prior beyond that of GR & $+5.3$ & 4.5 to 6.3 \\
    \addlinespace
    \multicolumn{3}{l}{\emph{\Cref{sec:s1}: model size at fixed data}} \\
    8B minus 4B at step 6,000 & $+0.77$ & $\pm0.37$ \\
    \addlinespace
    \multicolumn{3}{l}{\emph{\Cref{sec:forgetting}: forgetting rather than crowding out}} \\
    First-level gain, 32M minus 8M URLs at equal training after the pair, pooled over five models (nats) & $+0.12$ & $\pm0.07$ \\
    \addlinespace
    \multicolumn{3}{l}{\emph{\Cref{sec:window}: the decay window}} \\
    End of round 3, pairs drawn inside minus before the window & $+15.3$ & 11.7 to 19.3 \\
    \quad the same, pairs drawn by clicks & $+1.8$ & $-2.1$ to $+6.3$ \\
    End of round 4, pairs drawn inside minus before the window & $-1.0$ & $-8.0$ to $+6.3$ \\
    Round 4, released minus constant-rate arm, early in the window (\trafficjul) & $+5.9$ & 4.3 to 7.5 \\
    \addlinespace
    \multicolumn{3}{l}{\emph{\Cref{sec:batch}: batch size}} \\
    4B grid model, twice the batch size minus the original batch size (trained pairs) & $+5.6$ & $\pm0.5$ \\
    \addlinespace
    \multicolumn{3}{l}{\emph{\Crefrange{sec:expo}{sec:spread}: more exposures; repeated against spread exposures}} \\
    0.6B at 16 exposures minus 14B at 2, trained pairs & $+3.1$ & $\pm0.7$ \\
    0.6B, held-out queries, 16 minus 2 exposures & $-3.5$ & $\pm1.2$ \\
    0.6B, held-out queries, 16 exposures minus the starting point & $+0.8$ & $\pm1.2$ \\
    Held-out queries at 2 exposures, 14B minus 0.6B & $+4.1$ & $\pm1.3$ \\
    Spread minus repeated exposures, held-out queries & $+12.0$ & $\pm1.7$ \\
    \quad the same, queries similar to a training query of their URL & $+13.5$ & $\pm2.1$ \\
    \quad the same, queries similar to none & $+7.3$ & $\pm2.8$ \\
    \bottomrule
  \end{tabular}
\end{table}

\subsection{Weighting every query once}
\label{app:macro}

Weighting by clicks, as hit@$k$ does (\cref{eq:hitk}), measures how often a user gets the right page. Weighting every navigational query
once instead measures how many distinct queries a system handles, and it gives the long tail of rare queries far more weight
(\cref{tab:macro}). To compute it, we reweight each sampled query by the inverse of its probability of entering the sample; queries with
fewer than 1,000 clicks enter through a 2\% hash sample before the draws. This reconstructs about 83M navigational queries with at least
10 September clicks. Of these, 39\% have fewer than 100 clicks but carry only 6\% of the traffic, and 1.6\% have at least 1,000 clicks and
carry 56\% of it. Because the tail dominates the count, every system is far weaker per query. GR remains ahead of dense retrieval by 7.0
points of hit@1 (95\% interval 5.7 to 8.4) and of BM25 by 4.0, and it is level with the lookup ($-0.1$, $-1.5$ to $+1.4$). \modelsmall{}
falls below dense retrieval per query, because its smaller universe misses much of the tail.

\begin{table}[H]
  \centering
  \small
  \caption{\textbf{Navigational hit@1 / hit@10 (\%) on \trafficsep{}} weighted by clicks, as everywhere else in the report, and with
  every navigational query weighted once. $^\dagger$As in \cref{tab:main}. $^\ddagger$Returns URLs from \Usmall.}
  \label{tab:macro}
  \begin{tabular}{lcc}
\toprule
 & Weighted by clicks & Each query once \\
\midrule
Octen-WebGR-4B & \textbf{59.6} / \textbf{68.7} & \textbf{22.0} / \textbf{34.8} \\
Octen-WebGR-0.6B$^\ddagger$ & 56.5 / 64.0 & 14.3 / 21.9 \\
Dense retrieval (exact) & 24.8 / 55.6 & 15.0 / 32.2 \\
BM25 (URL + queries) & 9.1 / 25.3 & 18.0 / 30.5 \\
Click-table lookup$^\dagger$ & 58.9 / 62.5 & 22.1 / 24.2 \\
\midrule
Lookup $\rightarrow$ GR & \textbf{62.5} / \textbf{70.2} & \textbf{29.2} / \textbf{39.5} \\
Lookup $\rightarrow$ dense & 61.8 / 70.1 & 27.7 / 38.9 \\
\bottomrule
\end{tabular}

\end{table}

\subsection{New queries by the familiarity of the target}
\label{app:newq}

This breakdown shows the crossover that separates familiar from cold targets. \Cref{tab:newq} divides the new queries of \trafficsep{} by
the number of training pairs of their target; the familiar and cold blocks of \cref{tab:main} merge its buckets above and below ten
pairs. GR's hit@1 rises steadily with the target's training pairs, from 6.4\% with one pair to 35.3\% with more than 50. Dense retrieval
and BM25 instead peak at 4 to 10 pairs and fall beyond, since a page with many queries shares its keywords with many other pages.

\begin{table}[H]
  \centering
  \small
  \caption{\textbf{New queries of \trafficsep{} by the number of training pairs of the target URL}: navigational hit@1 / hit@10 (\%),
  with each bucket's share of new-query traffic. A target outside \Ufull{} is a miss for every system. Bold: best system in each
  row.}
  \label{tab:newq}
  \begin{tabular}{lrccc}
\toprule
Training pairs of the target URL & Traffic & Octen-WebGR-4B & Dense retrieval (exact) & BM25 \\
\midrule
not in $\mathcal{U}_{\text{201M}}$ & 56.1\% & 0.0 / 0.0 & 0.0 / 0.0 & 0.0 / 0.0 \\
1 pair & 2.1\% & 6.4 / 17.1 & 13.6 / 33.6 & \textbf{19.3} / \textbf{35.7} \\
2--3 pairs & 5.1\% & 14.0 / 32.0 & 22.7 / \textbf{52.6} & \textbf{26.2} / 48.0 \\
4--10 pairs & 9.5\% & 16.2 / 35.8 & \textbf{27.5} / \textbf{56.3} & 26.6 / 53.9 \\
11--50 pairs & 12.9\% & \textbf{26.2} / \textbf{52.7} & 20.3 / 50.6 & 16.2 / 41.9 \\
$>$50 pairs & 14.2\% & \textbf{35.3} / \textbf{63.3} & 14.5 / 41.6 & 2.7 / 14.1 \\
\midrule
All new queries & 100\% & \textbf{10.8} / \textbf{21.2} & 8.7 / \textbf{21.2} & 6.8 / 15.8 \\
\bottomrule
\end{tabular}

\end{table}

\subsection{Construction of \heldnav}
\label{app:heldnav}

\heldnav{} asks how well a system answers a new query for a URL that is in the universe. We describe how its queries were withheld and
then the four channels through which a withheld query could still reach a system, each of which we close or measure.

\heldnav{} draws navigational queries from the training data: each has at least ten clicks, at least half of which fall on one URL, its
target. It takes 20,000 queries from each of three bands of query clicks (at least 1,000, 100 to 999, and fewer than 100) and splits them
into a development and a test half by a hash of the query. Together with a sample of other queries and the queries of an earlier
evaluation set, the 150,614 withheld queries were removed from SID training with their pairs. On \heldnav{} the lookup therefore has no
entry for them, whereas on \trafficsep{} the lookup, which uses all clicks up to August, still answers them (the withheld block of
\cref{tab:main}). Of the 29,894 test queries, 24,519 have their target in \Ufull; the shares below are of these.

\paragraph{URL texts.} The text of a URL, its normalised form followed by its five most-clicked queries, is what BM25 indexes, what the
dense retriever encodes and what the SIDs are built from. For 77.0\% of the queries (83\% in the head band, 64\% in the tail), the query
itself is one of these five for its target. We therefore rebuilt the texts of the 460,266 URLs of \Ufull{} that list a withheld query:
the withheld queries are removed, and the URL's next most-clicked queries take their places up to the original number (181,421 URLs have
no other query and keep fewer). The dense retriever re-encodes these URLs, BM25 indexes the cleaned texts, and the click prior subtracts
the withheld queries' clicks. The cleaning matters: with the original texts, dense retrieval and BM25 rank the target first for 35.7\%
and 45.3\% of the queries, against 20.6\% and 16.5\% with the cleaned texts (\cref{tab:heldnavleak}).

\paragraph{July data.} Two components saw July data: GR's URL-string pre-training taught it to generate URL strings for the queries and keywords of July 2026
(\cref{sec:rounds}), and the dense retriever's maps were trained on July navigational queries. For 44.4\% of the queries, the query is a
keyword of its target in the July data, and 10.3\% are among the dense retriever's training queries, 47.1\% in all. We exclude these,
which leaves 12,967 queries. Popular queries are the most likely to recur, so the exclusion removes most of the head band: 1,881 of its
9,922 queries remain, against 5,981 of 8,267 in the torso and 5,105 of 6,330 in the tail. The new queries of \trafficsep{} are almost
untouched by this channel, since only 0.4\% of their traffic occurs in the July data.

\paragraph{Identifiers.} The SIDs were built from the uncleaned texts and cannot be rebuilt without retraining, so we measure instead
whether a withheld query moved its target's identifier. For each affected target we compare the nearest first-level centroid of the
embedding of its original text with that of its cleaned text; the former reproduces the assigned first code for 99.4\% of the targets.
The two differ for 34.3\% of the affected targets of the evaluated queries. If the identifiers carried the query, GR should gain on these
targets; instead it is least accurate on them (13.7\% hit@1, against 22.7\% where the first code is kept; \cref{tab:heldnavleak}). We see
no sign of this channel, though a smaller effect cannot be excluded. Where the query is not among its target's keywords, so that none of
these channels is open, the target has at least five more-clicked queries, and GR ranks it first for 54.0\% of the queries against
19.2\% for dense retrieval.

\paragraph{Indexing samples.} GR's document-indexing samples map the uncleaned text of a URL to its identifier (\cref{sec:samples}), so for
most targets GR has been trained on a text that contains the withheld query. How much of that training remains differs between old and
new pages. Old pages, 11,919 of the 12,967 targets, received such samples only in the round on \Ucore. Training after that round
accumulated $\Lambda\approx8.6$, far above the 0.56 beyond which trained pairs are forgotten (\cref{sec:window}). Consistently, at equal
numbers of training pairs of the target (11--20, 21--50 and 51--100), GR is as accurate when the query is in the target's text as when it
is not (27.7\%, 36.7\% and 59.8\% against 28.4\%, 36.8\% and 60.2\% hit@1), whereas a lasting effect of the samples would favour the
former. New pages received indexing samples in a random quarter or half of each later round, and 99.6\% of the 1,048 new-page targets
have the query in their text. GR ranks those indexed in rounds 3 or 4 (466) first $4.4\pm4.4$ points more often than the others; this
includes the general benefit of indexing new pages (\cref{app:newpages}), and dense retrieval, which no indexing sample reaches, differs by
$-3.5\pm4.7$ points between the two groups. Excluding the indexed targets raises GR's hit@1 on \heldnav{} from 32.0\% to 32.5\%, so they do
not inflate it.

\paragraph{Training pairs of the target.} \Cref{tab:heldnavpairs} divides \heldnav{} by the training pairs of the target, as \cref{tab:newq}
divides the new queries of \trafficsep. Dense retrieval and BM25 lead up to ten pairs and GR from eleven, the same crossover as in
traffic. Unlike traffic, \heldnav{} has targets without any training pair: for 1,366 queries the withheld pairs were all that the target
had, and on these GR ranks the target first for 6.3\% of the queries and dense retrieval for 10.1\%.

\begin{table}[H]
  \centering
  \small
  \caption{\textbf{\heldnav{} by the number of training pairs of the target URL}: hit@1 / hit@10 (\%) on its 12,967 queries, with the
  number of queries per bucket. Withheld pairs are not counted. Bold: best per row.}
  \label{tab:heldnavpairs}
  \begin{tabular}{lrccc}
\toprule
Training pairs of the target URL & Queries & Octen-WebGR-4B & Dense retrieval (exact) & BM25 \\
\midrule
0 pairs & 1,366 & 6.3 / 19.6 & 10.1 / 21.4 & \textbf{13.3} / \textbf{23.1} \\
1 pair & 1,252 & 12.7 / 30.6 & 16.0 / 35.5 & \textbf{22.9} / \textbf{43.1} \\
2--3 pairs & 1,624 & 16.4 / 36.8 & 27.3 / \textbf{55.1} & \textbf{29.9} / 54.7 \\
4--10 pairs & 2,212 & 18.5 / 43.0 & 28.2 / 59.4 & \textbf{30.8} / \textbf{60.1} \\
11--50 pairs & 2,514 & \textbf{32.6} / \textbf{59.5} & 19.3 / 53.7 & 14.4 / 46.3 \\
$>$50 pairs & 3,999 & \textbf{60.2} / \textbf{83.4} & 19.6 / 53.0 & 3.6 / 21.8 \\
\bottomrule
\end{tabular}

\end{table}

\begin{table}[H]
  \centering
  \small
  \setlength{\tabcolsep}{3pt}
  \caption{\textbf{\heldnav{} by the query's place in its target's URL text}: hit@1 / hit@10 (\%), with numbers of queries in
  parentheses. ``Not in text'': the query is not among the target's keywords; ``in text'': it is, and the first code of the target's
  identifier would be kept or would move if the identifier were built from the cleaned text. Dense retrieval and BM25 are shown with the
  original and the cleaned texts. $^\ddagger$Returns URLs from \Usmall. Bold: best per column among the systems that use the cleaned
  texts.}
  \label{tab:heldnavleak}
  \begin{tabular}{lcccc}
\toprule
 & All & Not in text & In text, $s_0$ kept & In text, $s_0$ moves \\
 & (12,967) & (4,658) & (5,458) & (2,851) \\
\midrule
Octen-WebGR-4B & \textbf{32.0} / \textbf{54.2} & \textbf{54.0} / 78.4 & 22.7 / 44.6 & 13.7 / 33.2 \\
Octen-WebGR-0.6B$^\ddagger$ & 27.5 / 42.7 & 53.7 / \textbf{78.9} & 15.0 / 25.3 & 8.6 / 17.1 \\
Dense retrieval, cleaned text & 20.6 / 49.5 & 19.2 / 53.0 & 23.5 / 51.5 & 17.4 / \textbf{39.9} \\
Dense retrieval, original text & 35.7 / 68.5 & 18.2 / 52.5 & 43.3 / 76.6 & 49.4 / 79.2 \\
BM25, cleaned text & 16.5 / 39.4 & 5.3 / 25.4 & \textbf{25.2} / \textbf{51.9} & \textbf{18.4} / 38.2 \\
BM25, original text & 45.3 / 70.6 & 4.6 / 24.2 & 68.0 / 96.6 & 68.2 / 96.5 \\
\bottomrule
\end{tabular}

\end{table}

\subsection{Dense encoders}
\label{app:dense}

Public encoders show how much of the dense baseline's behaviour comes from its training on clicks. We encoded the 32M-URL subset of
\cref{sec:ladder} with Qwen3-Embedding~\citep{zhang2025qwen3embedding} and Octen-Embedding~\citep{octen2025embedding} at 0.6B, 4B and 8B,
used as their model cards prescribe, and trained the linear maps of the dense baseline on the three Octen encoders with the recipe of one
of its runs. On \trafficsep{} over this subset, where the dense baseline reaches 37.0\% hit@1 and GR 62.7\%, all six encoders reach only
16--18\% zero-shot, regardless of size; the trained maps lift the Octen models to 34.9\%, 36.3\% and 37.9\%. On \heldnav{}
(\cref{tab:heldnavzs}), whose queries are drawn equally from the head, torso and tail rather than by clicks, the zero-shot encoders do far
better: about 21\% on its head band, close to their level on traffic, and 30--35\% on the torso and tail.

\begin{table}[H]
  \centering
  \small
  \setlength{\tabcolsep}{3.5pt}
  \caption{\textbf{Dense encoders on \heldnav}: hit@1 / hit@10 (\%) on its 12,967 queries over a 32.4M-URL subset, the 32M subset of
  \cref{sec:ladder} extended with every \heldnav{} target and the 50 most-clicked pages of its host; familiar targets have at least 11
  training pairs. ``$\cup$ GR'' is the share of queries whose target is in the union of the encoder's and GR's top ten.}
  \label{tab:heldnavzs}
  \begin{tabular}{lrcccc}
\toprule
Encoder & Dims & All & Familiar & Cold & $\cup$ GR \\
\midrule
Octen-WebGR-4B (GR) & -- & \textbf{39.8} / 62.5 & \textbf{57.3} / \textbf{79.7} & 22.1 / 45.1 & -- \\
Dense baseline & 1,024 & 34.8 / 65.1 & 34.4 / 70.2 & 35.2 / 60.1 & 78.9 \\
\midrule
\multicolumn{6}{l}{\emph{Zero-shot}} \\
Qwen3-Embedding-0.6B & 1,024 & 29.2 / 57.3 & 19.2 / 51.3 & 39.2 / 63.4 & 79.4 \\
Qwen3-Embedding-4B & 2,560 & 31.0 / 61.9 & 19.0 / 54.9 & 43.1 / 68.9 & 81.4 \\
Qwen3-Embedding-8B & 4,096 & 32.4 / 64.5 & 20.0 / 58.4 & \textbf{44.9} / \textbf{70.7} & \textbf{82.1} \\
Octen-Embedding-0.6B & 1,024 & 29.2 / 57.3 & 19.3 / 51.9 & 39.1 / 62.8 & 79.0 \\
Octen-Embedding-4B & 2,560 & 30.4 / 60.9 & 19.2 / 55.9 & 41.6 / 66.0 & 80.2 \\
Octen-Embedding-8B & 4,096 & 32.4 / 63.5 & 20.9 / 58.4 & 44.0 / 68.7 & 81.4 \\
\midrule
\multicolumn{6}{l}{\emph{With linear maps trained on navigational pairs}} \\
Octen-Embedding-0.6B & 1,024 & 29.7 / 58.5 & 30.7 / 66.7 & 28.7 / 50.2 & 76.3 \\
Octen-Embedding-4B & 1,024 & 32.2 / 63.0 & 32.7 / 69.4 & 31.7 / 56.6 & 78.1 \\
Octen-Embedding-8B & 1,024 & 35.4 / \textbf{66.2} & 34.7 / 71.1 & 36.2 / 61.3 & 79.4 \\
\bottomrule
\end{tabular}

\end{table}

\subsection{Dense baseline}
\label{app:densecfg}

The dense baseline of \cref{sec:baselines} embeds queries and URLs with the frozen Octen-Embedding-4B encoder, truncates the embeddings to
their first 1,024 dimensions and passes them through one linear map per side, $W = I + \Delta W$ with $\Delta W$ initialised at zero, whose
outputs are renormalised. Only the maps, 2.1M parameters, are trained, with a contrastive loss (temperature 0.02) on navigational
query--URL pairs from the click data, in batches of 16,384 queries for 12,000 steps. The final maps average seven such runs that differ in
their negatives (taken from the click data, mined from a search of the full index or shared across the batch), learning rate and
starting point, with the averaged $\Delta W$ scaled by 0.95. Training, including the mining, took about 23 H200 GPU-hours. The click prior's
weight, 0.04, was chosen on \trafficjul.

\subsection{Near misses of dense retrieval}
\label{app:denseerr}

For 30.8\% of the navigational traffic of \trafficsep, dense retrieval ranks the target second to tenth. \Cref{tab:denseerr} sorts this
traffic by how dense retrieval's first result relates to the target. The kinds are assigned by rules on the two URL strings, which use a
list of encyclopedia hosts and, to find registered domains, a list of two-part suffixes such as \texttt{co.uk}. The rules apply in this
order: an encyclopedia host for the first result only; the same URL once language and region markers such as \texttt{fr-ca} or
\texttt{hl=es} are removed; the same name under another domain ending; a login, sign-up or account page of the target's registered
domain; another page of that domain; a similar name, equal or contained in the other once hyphens and digits are removed, or within one
or two edits; and any other site. Since targets are defined by clicks, some first results of the kinds ``another language or region''
and ``same name, other domain ending'' may suit some users as well, such as \texttt{bbc.com} for users outside the United Kingdom.

\begin{table}[H]
  \centering
  \footnotesize
  \setlength{\tabcolsep}{3pt}
  \caption{\textbf{What dense retrieval ranks first when the target is second to tenth}, as shares of this traffic, by whether the first
  result is on the target's site, with one example of each kind among frequent queries. GR ranks the target first for 81.5\% of this
  traffic and for every example.}
  \label{tab:denseerr}
  \begin{tabular}{lrlll}
\toprule
 & & \multicolumn{3}{l}{Example} \\
\cmidrule(l){3-5}
Dense retrieval's first result & \% & Query & Target & First result \\
\midrule
\emph{On the target's site} & 46.1 & & & \\
\quad another page & 34.7 & \texttt{whatsapp web} & \texttt{web.whatsapp.com} & \texttt{faq.whatsapp.com/1317564962315842} \\
\quad another language or region & 7.6 & \texttt{chatgpt} & \texttt{chatgpt.com} & \texttt{chatgpt.com/fr-ca} \\
\quad login or sign-up page & 3.8 & \texttt{canva} & \texttt{canva.com} & \texttt{canva.com/login} \\
\emph{On another site} & 53.9 & & & \\
\quad same name, other domain ending & 6.4 & \texttt{bbc news} & \texttt{bbc.co.uk/news} & \texttt{bbc.com} \\
\quad look-alike domain & 5.7 & \texttt{threads} & \texttt{threads.com} & \texttt{thread.com} \\
\quad encyclopedia article & 7.1 & \texttt{epic games} & \texttt{store.epicgames.com} & \texttt{en.wikipedia.org/wiki/epic\_games} \\
\quad any other site & 34.6 & \texttt{wa web} & \texttt{web.whatsapp.com} & \texttt{wa.gov.au} \\
\bottomrule
\end{tabular}

\end{table}

\subsection{Serving cost}
\label{app:serving}

\Cref{tab:serving} gives the measurements and assumptions behind \cref{fig:cost} and \cref{sec:serving}.

\begin{table}[H]
  \centering
  \small
  \setlength{\tabcolsep}{3pt}
  \caption{\textbf{Serving cost per million queries.} GPU time is measured on H200 GPUs: for GR with our decoding code (batches of 64
  queries per GPU, the prompt's key--value cache reused), for dense retrieval as query encoding plus exact search, the latter scaled
  linearly from 5,200 queries per second measured over 17M vectors on one GPU. The CPU row gives socket-milliseconds and socket-hours
  per query in batches of 8 (57 queries per second on one socket; a single query takes 72\,ms). Dollars assume \$3 per H200-hour. hit@1 is on
  \trafficsep{}; for the LLMs it is on the 5,000-query sample of \cref{tab:llm}, on which \model{} reaches 59.8\%. LLM costs assume 20
  input and 170 output tokens per query at list prices (October 2026) and count no reasoning tokens, so they are lower bounds.
  Self-hosted costs count measured compute only, with no provider margin, whereas list prices include it. In \cref{fig:cost}, two LLMs
  with the same price are drawn slightly apart.
  $^\ast$Prices from third-party listings of OpenAI's API, not checked against OpenAI's own page; for Sol, the price listed for GPT-6.1
  Sol.}
  \label{tab:serving}
  \begin{tabular}{lrrrlc}
\toprule
System & GPU-ms / query & H200-h / $10^6$ & \$ / $10^6$ & Resident & hit@1 \\
\midrule
Octen-WebGR-4B & 2.3 & 0.65 & 1.9 & 9.7\,GiB, one GPU & \textbf{59.6} \\
Octen-WebGR-0.6B & \textbf{1.6} & \textbf{0.45} & \textbf{1.4} & 1.6\,GiB, one GPU & 56.5 \\
Dense (encoder + exact search) & 3.9 & 1.08 & 3.2 & 8\,GB + 412\,GB, $\geq$3 GPUs & 24.8 \\
Octen-WebGR-0.6B, one CPU socket & 17.5 (CPU) & 4.9 (CPU) & -- & 1.6\,GiB & 56.5 \\
\midrule
\multicolumn{6}{l}{\emph{Frontier LLMs from memory, list prices, no reasoning tokens counted}} \\
Claude Haiku 5.5 & -- & -- & $\geq$87 & API & 29.1 \\
Claude Sonnet 5.5 & -- & -- & $\geq$1,740 & API & 34.5 \\
Claude Opus 5.5 & -- & -- & $\geq$3,480 & API & 36.8 \\
GPT-6 Luna$^\ast$ & -- & -- & $\geq$87 & API & 30.8 \\
GPT-6 Sol$^\ast$ & -- & -- & $\geq$1,740 & API & 35.1 \\
GPT-6 Astra$^\ast$ & -- & -- & $\geq$8,700 & API & 37.5 \\
\bottomrule
\end{tabular}

\end{table}

\section{Closed-Book Evaluation of LLMs}
\label{app:llm}

This appendix describes how the frontier LLMs of \cref{sec:llmres} were evaluated. Six models (GPT-6 Astra, Sol and Luna; Claude Opus 5.5,
Sonnet 5.5 and Haiku 5.5) gave ten URLs for each of 5,000 navigational queries sampled from \trafficsep{} by traffic, from memory and
without tools (\cref{tab:llm}). The two strongest, Astra and Opus, are within noise of each other (a paired 95\% confidence interval of
$-0.1$ to $+1.6$ points of hit@1). Many of the URLs the LLMs write are not pages that users clicked: only 38--61\% of them appear among the
349M clicked URLs of our pool. We do not map the LLMs' URLs onto the table, since a mapping that keeps each URL's host cannot exceed the
site-level column of \cref{tab:llm}, where Astra reaches 52.4\%, still below GR's 59.8\% for the exact URL.

\begin{table}[H]
  \centering
  \small
  \setlength{\tabcolsep}{3.2pt}
  \caption{\textbf{Closed-book LLMs against GR on the same 5,000 navigational \trafficsep{} queries} (hit@1 / hit@10, \%). ``Site''
  counts a URL on the target's host as correct. $^\dagger$As in \cref{tab:main}. $^\ddagger$Returns URLs from \Usmall. Bold: best per
  column among our systems and baselines and, separately, among the LLMs.}
  \label{tab:llm}
  \begin{tabular}{lcc}
\toprule
System & exact URL & site \\
\midrule
Octen-WebGR-4B & \textbf{59.8} / \textbf{68.4} & \textbf{67.6} / \textbf{80.7} \\
Octen-WebGR-0.6B$^\ddagger$ & 56.7 / 64.0 & 63.7 / 75.4 \\
Click-table lookup$^\dagger$ & 58.9 / 62.3 & 60.7 / 63.3 \\
Dense retrieval (exact) & 25.1 / 55.5 & 39.5 / 70.9 \\
\midrule
GPT-6 Astra & \textbf{37.5} / \textbf{49.5} & 52.4 / 70.4 \\
GPT-6 Sol & 35.1 / 43.9 & 49.3 / 66.3 \\
GPT-6 Luna & 30.8 / 39.5 & 44.2 / 60.8 \\
Claude Opus 5.5 & 36.8 / 46.3 & \textbf{52.7} / \textbf{70.7} \\
Claude Sonnet 5.5 & 34.5 / 43.9 & 50.1 / 66.9 \\
Claude Haiku 5.5 & 29.1 / 37.1 & 43.0 / 58.2 \\
\bottomrule
\end{tabular}

\end{table}

\paragraph{Protocol.} The queries were split into 20 batches of 250, and every batch was a separate session of the model in an agent
harness whose instruction file read:

\begin{scriptsize}
\begin{verbatim}
You are being evaluated as a closed-book web search engine. For each search
query below, predict the web pages (URLs) the user is most likely to click on,
using ONLY your own knowledge.

Strict rules:
- Do NOT look anything up. No web search, no web fetch, no shell commands, and
  do not open any file other than this instruction file. Answer purely from
  memory.
- The only tools you may use are Read (for this instruction file only) and
  Write (to save your answers).

For every query give exactly 10 distinct URLs, most likely first:
- Each URL must be complete and concrete (e.g. https://www.example.com/some/page):
  the actual page the user lands on (official site, login page, specific
  article, profile, product or video page), not a search-results page unless
  that really is what users click.
- The queries come from real search logs worldwide (September 2026), in many
  languages. When the query implies a country or language, give that local
  version of the site.
- Same effort for every query, first to last. Never skip a query and never
  write fewer than 10 URLs.

Write the answers in order, about 20 queries per file, ... in exactly this
format:
### <id> | <the query, copied exactly from the list>
<url 1>
...
<url 10>
\end{verbatim}
\end{scriptsize}

The GPT-6 models (reasoning effort \emph{medium}) ran in a harness with file and shell tools; their instruction added that the tools may
be used only to read the instruction and write the answers, and that a declined query may be left blank. We set no temperature or other
sampling parameter, so every model ran with its harness's defaults. Sessions that stopped early were resubmitted, and blocks with fewer
than ten valid URLs once more. Tool logs show no Claude batch searched, fetched pages or ran commands; we did not retain logs for the
GPT-6 runs. Each answer block must repeat its query, which guards against answers shifting to the wrong query, and URLs are normalised as
in \cref{sec:clickdata}.

Unanswered or declined queries, mostly about piracy or adult content, count as misses: 3 for Haiku, none for Sonnet, 15 for Opus (one
batch flagged by a safety filter) and 121, 2 and 12 for Astra, Sol and Luna. They do not change the comparison: restricted to the 4,864
queries that all six answered, no difference between systems moves by more than 0.6 points.

\section{Public Benchmark: ORCAS-Nav}
\label{app:public}

Our click data are proprietary, so we also evaluate on a benchmark built from public clicks. ORCAS~\citep{craswell2020orcas} releases
18.8M query--document clicks from Bing logs up to January 2020, restricted to US English and to the 3.2M documents of the TREC Deep
Learning corpus; ORCAS-I~\citep{alexander2022orcasi} labels 2.7M of them navigational. We sample 20,000 navigational queries with a fixed
seed, every query weighted equally (\cref{tab:orcas}). A target of 62.6\% of them is in \Ufull. Since the target documents were collected
years before our click data, the host-level match is the more informative of the two measures.

GR leads dense retrieval at hit@1 (16.2\% against 11.8\%; $+4.4$ points, 95\% confidence interval $+3.9$ to $+4.9$) and trails it by 1.3
points at hit@10. With the click prior, dense retrieval leads GR at both cutoffs (21.0\% and 39.9\%), whereas the same prior leaves GR's
hit@1 at 16.2\%. ORCAS-I labels a query navigational mainly when it resembles the clicked domain, so these queries are mostly site names,
and GR does as well on those absent from our training data as on those present in it.

\begin{table}[H]
  \centering
  \small
  \setlength{\tabcolsep}{2.1pt}
  \caption{\textbf{ORCAS-Nav: hit@1 / hit@10 (\%) for 20,000 navigational queries from public Bing clicks}, by overlap with the training
  data. ``Target in \Ufull'' is the share of queries with a target in the universe; the host-level row counts any URL on a target's host
  as correct. Dense + click prior and $^\dagger$as in \cref{tab:main}. $^\ddagger$Returns URLs from \Usmall. Bold: best system in each
  row.}
  \label{tab:orcas}
  \footnotesize
  \begin{tabular}{>{\raggedright\arraybackslash}p{2.9cm}rrcccccc}
\toprule
Stratum & Queries & \twoline{Target in}{$\mathcal{U}_{\text{201M}}$} & \twoline{Octen-WebGR-}{4B} & \twoline{Octen-WebGR-}{0.6B$^\ddagger$} & \twoline{Dense retrieval}{(exact)} & \twoline{Dense +}{click prior} & BM25 & \twoline{Click-table}{lookup$^\dagger$} \\
\midrule
All & 20,000 & 62.6\% & 16.2 / 30.5 & 17.4 / 33.3 & 11.8 / 31.8 & \textbf{21.0} / \textbf{39.9} & 2.3 / 8.8 & 6.0 / 10.2 \\
Query not a training query & 11,784 & 62.7\% & 16.4 / 30.4 & 17.8 / 33.6 & 12.9 / 33.1 & \textbf{22.4} / \textbf{40.7} & 2.0 / 7.3 & 0.0 / 0.0 \\
Query is a training query & 8,216 & 62.5\% & 15.8 / 30.7 & 16.9 / 32.9 & 10.1 / 30.0 & \textbf{19.0} / \textbf{38.7} & 2.7 / 10.9 & 14.5 / 24.9 \\
(query, target) is a training pair & 2,280 & 99.6\% & 46.1 / 73.3 & \textbf{46.8} / 74.8 & 27.3 / 68.5 & 42.3 / \textbf{78.3} & 9.0 / 32.8 & 52.3 / 89.8 \\
\midrule
All, target in $\mathcal{U}_{\text{201M}}$ & 12,517 & 100\% & 25.8 / 48.8 & -- & 18.8 / 50.9 & \textbf{33.6} / \textbf{63.7} & 3.6 / 14.1 & 9.5 / 16.4 \\
All, host-level match & 20,000 & -- & 33.0 / 53.2 & 34.0 / 53.9 & 29.7 / 55.2 & \textbf{36.2} / \textbf{58.1} & 12.1 / 29.9 & 13.3 / 19.3 \\
\bottomrule
\end{tabular}

\end{table}

\section{Scaling Experiments}
\label{app:ladder}

This appendix gives the details of the scaling experiments of \cref{sec:scaling}.

\subsection{Number of candidate URLs}

Every subset of \cref{sec:ladder} contains the same 291,453 URLs: the 11,178 navigational targets of \trafficsep{} and up to 50 of the
most-clicked pages of each of their 8,044 hosts. Only the number of other URLs therefore grows, as is common practice for such
ladders~\citep{froebe2025subsampling,wang2026bm25scale}.

\subsection{Model size at fixed data}
\label{app:s1}

\Cref{tab:s1sep} gives the runs of \cref{sec:s1} along their 6,000 steps. Larger models learn more from a single exposure: at step 6,000
the 4B model ranks first 37.2\% of the \trafficjul{} query--URL pairs it saw once in training and 68.7\% of those it saw twice.

\begin{table}[H]
  \centering
  \small
  \caption{\textbf{Model size at fixed data.} Navigational hit@1 / hit@10 (\%) on \trafficsep{} along 6,000 steps of training on the
  same samples over \Ufull, and hit@1 at step 6,000 on seen pairs and new queries.}
  \label{tab:s1sep}
  \begin{tabular}{lrrrrrr}
\toprule
Model & step 1500 & step 3000 & step 4500 & step 6000 & \multicolumn{2}{c}{step 6000, hit@1} \\
\cmidrule(lr){6-7}
 & & & & & seen pair & new query \\
\midrule
0.6B & 22.1 / 38.6 & 31.1 / 47.1 & 35.4 / 50.3 & 39.9 / 53.4 & 60.0 & 4.7 \\
1.7B & 25.1 / 41.2 & 33.8 / 49.9 & 38.9 / 53.2 & 43.1 / 56.7 & 65.0 & 5.4 \\
4B & 30.3 / 46.8 & 37.7 / 53.1 & 40.7 / 55.7 & 45.8 / 59.1 & 68.5 & 6.5 \\
8B & \textbf{31.1} / \textbf{47.4} & \textbf{38.7} / \textbf{54.3} & \textbf{41.6} / \textbf{56.9} & \textbf{46.5} / \textbf{59.9} & \textbf{69.8} & \textbf{6.6} \\
\bottomrule
\end{tabular}

\end{table}

\subsection{The fixed-exposure grid}
\label{app:grid}

The subsets are drawn among URLs with at least two training pairs. The starting point is trained for 1,764 steps at a peak learning rate of $10^{-4}$,
since a run on 2M URLs alone is too short to learn the task. A step packs 262,144 tokens, as in \cref{sec:objective}; the grid's samples
are shorter on average, so a step holds about 18,100 of them rather than 16,000. The test pairs are 20,000 primary pairs of the 2M subset
whose queries occur only once among the training pairs of the 32M subset. Likelihoods use the softmax normalised over the trie, and we
report them alongside hit@1 because continuous measures reveal trends that thresholded accuracy can hide~\citep{schaeffer2023mirage}.

Part of the decrease in accuracy reflects a harder task: the zero-shot NLL of the gold SID rises by 0.8--1.1 nats per fourfold increase in
URLs, all of it at the second identifier level (\cref{tab:s2full}). The rule that four times the URLs need about 2.5 times the parameters
comes from fits over the 15 runs. Fitting logit hit@1, or hit@1 itself, to the logarithms of the number of URLs and of the number of
parameters gives 2.61 and 2.49 times the parameters for four times the URLs (2.3 to 2.9 when any one run is left out), or about ten times
the training compute, since the samples grow with the URLs; logit hit@10 gives 2.4. Fitted to the NLL reduction per pair instead, the rule
gives 1.0, because what a pair learns barely depends on the number of URLs.

\begin{table}[H]
  \centering
  \small
  \setlength{\tabcolsep}{4pt}
  \caption{\textbf{Every run of the fixed-exposure grid}, on the same 20,000 trained pairs. NLL is the trie-normalised negative
  log-likelihood of the gold SID; $\Delta$NLL is the zero-shot NLL minus the trained NLL, in total and per identifier level. Every run
  uses the constant learning rate of $3.5\times10^{-5}$ of \cref{sec:grid}. Bold: best on each subset.}
  \label{tab:s2full}
  \begin{tabular}{llrrrrrr}
\toprule
Model & Subset & \multicolumn{2}{c}{hit@1 / hit@10} & \multicolumn{2}{c}{NLL (nats)} & \multicolumn{2}{c}{$\Delta$NLL} \\
\cmidrule(lr){3-4}\cmidrule(lr){5-6}\cmidrule(lr){7-8}
 & & trained & zero-shot & trained & zero-shot & total & first / second level \\
\midrule
0.6B & 2M & 40.2 / 66.0 & 23.6 / 45.9 & 3.95 & 5.45 & 1.50 & 1.03 / 0.47 \\
0.6B & 8M & 31.1 / 56.6 & 18.4 / 39.3 & 4.76 & 6.28 & 1.52 & 0.92 / 0.60 \\
0.6B & 32M & 20.5 / 44.8 & 12.6 / 31.0 & 5.82 & 7.32 & 1.50 & 0.83 / 0.67 \\
1.7B & 2M & 50.0 / 74.0 & 24.2 / 47.1 & 3.31 & 5.30 & 1.99 & 1.39 / 0.61 \\
1.7B & 8M & 37.7 / 65.0 & 19.0 / 40.7 & 4.10 & 6.12 & 2.02 & 1.23 / 0.79 \\
1.7B & 32M & 24.3 / 51.0 & 12.6 / 32.3 & 5.24 & 7.16 & 1.91 & 1.03 / 0.88 \\
4B & 2M & 78.9 / 91.2 & 23.2 / 46.4 & 1.66 & 5.35 & 3.69 & 2.66 / 1.03 \\
4B & 8M & 64.0 / 84.1 & 18.3 / 39.9 & 2.45 & 6.17 & 3.72 & 2.31 / 1.41 \\
4B & 32M & 37.9 / 65.6 & 12.1 / 31.5 & 4.02 & 7.23 & 3.21 & 1.71 / 1.49 \\
8B & 2M & 87.1 / 95.1 & 21.8 / 44.7 & 1.11 & 5.49 & 4.38 & 3.20 / 1.19 \\
8B & 8M & 76.1 / 91.1 & 17.0 / 38.0 & 1.71 & 6.33 & 4.62 & 2.88 / 1.73 \\
8B & 32M & 49.2 / 74.6 & 11.4 / 30.4 & 3.25 & 7.43 & 4.18 & 2.22 / 1.96 \\
14B & 2M & \textbf{93.7} / \textbf{97.8} & 21.4 / 43.4 & 0.60 & 5.60 & \textbf{5.00} & 3.67 / 1.33 \\
14B & 8M & \textbf{87.5} / \textbf{95.9} & 16.3 / 37.3 & 1.00 & 6.46 & \textbf{5.45} & 3.44 / 2.01 \\
14B & 32M & \textbf{64.3} / \textbf{84.8} & 11.2 / 29.6 & 2.31 & 7.57 & \textbf{5.26} & 2.79 / 2.47 \\
\bottomrule
\end{tabular}

\end{table}

\subsection{Smaller universes}
\label{app:smalluniverse}

\Cref{tab:smalluniverse} gives the runs of \cref{sec:smalluniverse}. Because \Usmall{} shares the SIDs of \Ufull, every model trained on
one universe can also be decoded over the other. For comparison, the same 6,000-step runs on \Ucore{} reach 45.1\%, 48.1\%, 50.0\% and
50.9\% hit@1 at 0.6B, 1.7B, 4B and 8B. The sample mix explains part of the gain from a smaller universe: training on \Ufull{} with the
sample mix of \Usmall{} gives the 0.6B model 2.7 points (2.2 to 3.2) of the 10.2 it gains from the smaller universe.

\begin{table}[H]
  \centering
  \small
  \setlength{\tabcolsep}{4pt}
  \caption{\textbf{Navigational hit@1 / hit@10 (\%) on \trafficsep{} for models trained on \Ufull{} or on \Usmall{} and decoded over
  either}; a target outside the decoding universe counts as a miss. Bold: best per model.}
  \label{tab:smalluniverse}
  \begin{tabular}{lcccc}
\toprule
 & \multicolumn{2}{c}{Trained on $\mathcal{U}_{\text{201M}}$} & \multicolumn{2}{c}{Trained on $\mathcal{U}_{\text{31M}}$} \\
\cmidrule(lr){2-3}\cmidrule(lr){4-5}
Model & decoded over $\mathcal{U}_{\text{201M}}$ & over $\mathcal{U}_{\text{31M}}$ & decoded over $\mathcal{U}_{\text{201M}}$ & over $\mathcal{U}_{\text{31M}}$ \\
\midrule
0.6B, 6,000 steps & 39.9 / 53.4 & 40.3 / 53.6 & 43.8 / 58.1 & \textbf{50.1} / \textbf{60.3} \\
1.7B, 6,000 steps & 43.1 / 56.7 & 43.5 / 56.0 & 48.1 / 60.8 & \textbf{52.0} / \textbf{61.6} \\
4B, 6,000 steps & 45.8 / 59.1 & 45.8 / 57.9 & 48.9 / 62.3 & \textbf{53.3} / \textbf{62.9} \\
8B, 6,000 steps & \textbf{46.5} / \textbf{59.9} & 46.4 / 58.7 & -- & -- \\
\midrule
4B, end of round 3 & \textbf{58.8} / \textbf{68.1} & 57.0 / 64.0 & -- & -- \\
4B, round 4 (constant rate) & \textbf{59.2} / \textbf{68.7} & 56.9 / 64.1 & -- & -- \\
0.6B, 43.9K steps & \textbf{51.5} / \textbf{62.6} & 51.2 / 61.0 & -- & -- \\
Octen-WebGR-0.6B (43.9K steps) & -- & -- & 38.2 / 52.8 & \textbf{56.5} / \textbf{64.0} \\
\midrule
Targets in the universe & 79.2 & 68.9 & 79.2 & 68.9 \\
\bottomrule
\end{tabular}

\end{table}

\section{Forgetting and Exposure Analyses}
\label{app:limits}

This appendix gives the details of the analyses of \cref{sec:limits}: what the end points of the training rounds retain, forgetting in
the grid, the batch-size arms, the bits-per-parameter estimate, more exposures per query--URL pair, spread against repeated exposures, and new pages.

\subsection{What the end point retains}
\label{app:window}

\begin{table}[H]
  \centering
  \small
  \setlength{\tabcolsep}{4pt}
  \caption{\textbf{Rounds of SID training for \model{}} and the navigational hit@1 / hit@10 (\%) of each end point on \trafficsep.
  Rounds 3 and 4 close with a curated decay window; the constant-rate arm of round 4 tests the decay schedule (\cref{sec:schedule}) and
  shares its first 6,000 steps with the released arm.}
  \label{tab:rounds}
  \begin{tabular}{llr>{\raggedright\arraybackslash}p{7.4cm}c}
    \toprule
    Round & Universe & Steps & Decay window & \trafficsep \\
    \midrule
    \Ucore{} & \Ucore & 46K & last 10\% of steps, linear; not curated & -- \\
    1 & \Ufull & 27.3K & last 10\% of steps, linear; not curated & 54.1 / 65.7 \\
    2 & \Ufull & 26.3K & last 10\% of steps, linear; not curated & 54.8 / 65.6 \\
    3 & \Ufull & 25.9K & last 110.9M samples (27.5\%), linear decay: dominant pairs of frequent navigational queries and the end of the
      main sequence & 58.8 / 68.1 \\
    4 (released) & \Ufull & 22.1K & last 249.5M samples (74\%), exponential decay; same composition as in round 3 & \textbf{59.6} / \textbf{68.7} \\
    4, constant rate & \Ufull & 22.1K & $3.5\times10^{-5}$ from step 6,000, then a linear decay over the last 2,257 steps of the round & 59.2 / \textbf{68.7} \\
    \bottomrule
  \end{tabular}
\end{table}

\Cref{tab:rounds} lists the rounds of SID training. We measure retention on \trafficjul, whose targets are mostly training pairs, and
compare pairs with the same number of exposures. Only decayed checkpoints show what a run retains: the mid-round checkpoints of round 1 scored 57--61\% hit@1 on \trafficjul{} against 64.1\% at its end, which is why the recipe of \cref{sec:recipe} evaluates decayed checkpoints only. The
retention thresholds of \cref{sec:window} are measured at hit@10 against the cumulative learning rate $\Lambda$ after a pair's last
exposure in round 1. Steps alone do not predict forgetting: single-exposure pairs in the first fifth of the window of round 3 are
followed by about 5,900 steps but a cumulative learning rate of only about 0.30, and they end at 88.2\% hit@1, level with pairs seen
later.

The randomised comparison of \cref{sec:window} uses pairs that the curation left alone, drawn exactly once in round 3, inside against
before the window. On \trafficjul{} itself, measured against the end point of round 2, it has fewer pairs and wider intervals: $+8.3$
points ($-1.1$ to $+17.3$) for old pages and $+10.9$ ($+3.1$ to $+18.6$) for new pages. Pairs that the curation placed in the window are
not comparable with the rest: at the end of round 3, an old page's pair seen once inside the window is ranked first 87.0\% of the time and
one not seen in it 40.1\%, but the two groups already differed at the end of round 2, relative to which the difference is $+9.5$. The two
arms of round 4 (\cref{sec:schedule}) retain old-page pairs seen once in the window equally often (77.4\% and 77.3\%).

\subsection{Forgetting in the grid}
\label{app:forgetting}

In every run of the grid, the first-level gain decreases with the cumulative learning rate after a pair. In the 32M runs, pairs first
seen at $\Lambda=0.2$ before the end keep 20--29\% less than pairs first seen at 0.04. Rank one is more sensitive than the gain: the 4B
and 8B models lose half of what a pair adds at rank one after a cumulative learning rate of 0.04 to 0.12, well before the hit@10 threshold
of the full runs. The total gain per pair changes little with the number of URLs (\cref{tab:s2full}), because the second level, where more
URLs share each first-level code, gains 0.2 to 1.1 nats more from 2M to 32M URLs and offsets the first-level loss.

The matched comparison of \cref{sec:forgetting} needs both exposures of a pair fixed. Binning pairs by the learning rate accumulated after
their first exposure leaves the second exposure free, and binning by the last leaves the first free; binned by the last, the subsets
separate at small values, because in the longer run the first exposure can lie further back. The comparisons of \cref{tab:s2matched}
therefore restrict both runs to pairs whose two exposures fell in a common part of the constant-rate phase. From 8M to 32M URLs the change
is then $0.00$ to $+0.15$ nats for every model, and $0.00$ to $+0.13$ for secondary pairs. The comparison equalises the training after a pair
but not the training before it, which is longer in the larger run, so the small positive residual may come from that earlier training.
The 2M run shares no such window with the 32M run and only a short one with the 8M run, so we match 8M with 32M URLs only.

The grid stores too little to reveal a capacity limit: on 32M URLs its gains amount to 0.19 bits per parameter for the 0.6B model and to
under 0.1 for the 4B model, far below published capacity estimates. \Cref{app:expo} therefore raises the exposures of the 0.6B model.

\begin{table}[H]
  \centering
  \small
  \setlength{\tabcolsep}{4pt}
  \caption{\textbf{32M minus 8M URLs in the fixed-exposure grid}: gain per pair at the first identifier level, in nats. ``Raw'' compares
  the run means. For equal training after the pair, both runs are restricted to pairs whose two exposures were
  followed by 0.016 to 0.056 of accumulated learning rate within the constant-rate phase (8,476 primary pairs of the 8M run, 553 of the 32M
  run) and standardised to the 8M run's distribution over four bins ($\pm$ one standard error); secondary pairs are the URLs' other trained pairs.}
  \label{tab:s2matched}
  \begin{tabular}{lrrr}
\toprule
Model & Raw & \multicolumn{2}{c}{Equal training after the pair} \\
\cmidrule(lr){3-4}
  & primary pairs & primary pairs & secondary pairs \\
\midrule
0.6B & $-0.09$ & $+0.14\pm0.05$ & $+0.13\pm0.06$ \\
1.7B & $-0.19$ & $+0.14\pm0.06$ & $+0.12\pm0.07$ \\
4B & $-0.59$ & $+0.00\pm0.09$ & $+0.00\pm0.10$ \\
8B & $-0.67$ & $+0.09\pm0.11$ & $+0.04\pm0.11$ \\
14B & $-0.65$ & $+0.15\pm0.13$ & $+0.00\pm0.13$ \\
\bottomrule
\end{tabular}

\end{table}

\subsection{Batch size}
\label{app:batch}

The batch-size arms retrain the 0.6B grid model from its starting point on the same 32M-URL samples in the same order, with half to four
times the batch size, and the 4B model with twice the batch size; the learning rate is either kept or scaled in proportion
(\cref{fig:mech}b,c). A pair's position, and hence the training that follows it, is the same in every arm up to the grouping into steps.
With the learning rate scaled, forgetting is the same in every arm: the first-level gain of pairs first seen with 0.17--0.25 of cumulative
learning rate still to come is 0.67--0.72 of that of pairs seen at 0.03--0.04 for the 0.6B model, and 0.65--0.66 for the 4B model.
Without scaling, a larger batch ends below the original batch. With scaling, hit@1 on the trained pairs rises with the batch: for the
0.6B model from 20.5\% at the grid's batch to 22.9\% at four times it, and for the 4B model from 37.9\% to 43.5\% at twice it
($+5.6\pm0.5$).

\subsection{Bits per parameter}
\label{app:bits}

The estimate of \cref{sec:bits} credits a pair with what training added to it, not with the full cost of naming its URL. About 37\% of
the 1.13 billion training pairs have a query that sends at least half of its clicks to one URL. Charging each recalled pair the 27.6 bits
needed to name one of 201M URLs would overstate what training stored, since the starting point of the grid, which knows the task but not
these pairs, already ranks 5.3\% of them first. We therefore count the reduction in the NLL of the gold SID relative to the starting point:
6.2 bits per pair for the round-3 end point (33.3\% recalled at rank one) and 6.9 for the released end point (39.4\%). The reference does
not inflate the estimate: using instead the end point of the \Ucore{} round, which never trained on the URLs added later, as the
reference for pairs of those URLs gives 7.8 bits per pair against 6.5. Bits per parameter count every parameter, including the 131,072
identifier embeddings we add (4.36B for the 4B model, 0.73B for the 0.6B model); over non-embedding parameters they are a fifth higher for
the 4B model and two thirds higher for the 0.6B model.

\subsection{More exposures per pair}
\label{app:expo}

The 4-, 8- and 16-exposure runs start from the starting point of the 0.6B grid run on 32M URLs and replay its sample sequence in the
original order, so that each pass gives every pair two more exposures, at the same constant learning rate of $3.5\times10^{-5}$ with the
final 10\% of steps decayed. They share one constant-rate trunk: the 8-exposure run saved its state after 12,705 and 25,410 steps, from
which the 4-exposure arm decays and the 16-exposure arm continues. All arms are evaluated on the 20,000 trained pairs of the grid
(\cref{tab:s2expo}). The pairs can provide at most their zero-shot NLL,
7.32 and 7.68 nats for primary and secondary pairs, or 10.8 bits per pair: 0.95 bits per parameter (1.57 over non-embedding parameters), below
the 2 bits per parameter of published estimates. The ladder therefore shows that no limit lies below 0.69 bits per parameter, not where
one lies.

The held-out queries are navigational queries of the training period, each sending at least ten clicks and half of its clicks to one URL,
that were removed from SID training together with their pairs. We use 4,000 of the development half of the withheld queries of
\cref{app:heldnav} whose target is in the subset, the set on which the grid runs were monitored: the target's own pairs are trained, the
query is not. The arms share the identifiers and none starts from the URL-string pre-training of \model, so the channels of \cref{app:heldnav} through
which a withheld query can still reach a system affect them alike.

On these queries, size helps and repetition does not. The starting points, trained on no pair of the subset, already answer 11.2--13.9\% of
them at rank one, the larger models fewer. Two exposures of the subset's pairs then add $4.2\pm1.0$ points at 0.6B and $10.9\pm1.3$ at 14B,
whereas 16 exposures leave the 0.6B model only $0.8\pm1.2$ points above its starting point. Paired over the same queries, hit@1 changes by
$+0.4\pm0.8$, $-0.7\pm1.0$ and $-3.5\pm1.2$ points from two to 4, 8 and 16 exposures, and by $+4.1\pm1.3$ points from the 0.6B to the 14B
model at two exposures. The 16-exposure 0.6B model, trained in the same time as the 14B model, is $7.6\pm1.3$ points below it here and
$3.1\pm0.7$ above it on the trained pairs. FLOPs are $6ND$, with $N$ the non-embedding parameters and $D$ the 1.85B tokens of one pass over
the subset's samples, which gives each pair two exposures.

\begin{table}[H]
  \centering
  \footnotesize
  \setlength{\tabcolsep}{4pt}
  \caption{\textbf{The 32M-URL subset of the grid: the 0.6B model with 2 to 16 exposures per pair and the other models with two},
  after their starting points (0 exposures). Trained pairs are the grid's 20,000 test pairs; $\Delta$NLL is the gain per primary pair, as in
  \cref{tab:s2full}; bits per parameter count both pairs of every URL, over all parameters (over non-embedding parameters in
  parentheses); held-out queries are 4,000 navigational queries withheld from training whose target is in the subset.}
  \label{tab:s2expo}
  \begin{tabular}{lrrrcccc}
\toprule
Model & Exposures & FLOPs & Hours & Trained pairs & $\Delta$NLL & Bits per parameter & Held-out queries \\
 & per pair & & & hit@1 / hit@10 & (nats) & all (non-embedding) & hit@1 / hit@10 \\
\midrule
0.6B & 0 & -- & -- & 12.6 / 31.0 & -- & -- & 13.7 / 36.2 \\
0.6B & 2 & $4.9\times10^{18}$ & 1.8 & 20.5 / 44.8 & 1.50 & 0.19 (0.31) & 17.9 / 44.4 \\
0.6B & 4 & $9.8\times10^{18}$ & 3.6 & 26.2 / 53.0 & 2.27 & 0.29 (0.48) & 18.4 / 44.6 \\
0.6B & 8 & $2.0\times10^{19}$ & 7.2 & 41.1 / 69.0 & 3.63 & 0.46 (0.77) & 17.2 / 42.2 \\
0.6B & 16 & $3.9\times10^{19}$ & 14.4 & \textbf{67.4} / \textbf{88.1} & \textbf{5.39} & \textbf{0.69} (\textbf{1.15}) & 14.5 / 37.3 \\
\addlinespace
1.7B & 0 & -- & -- & 12.6 / 32.3 & -- & -- & 13.9 / 36.2 \\
1.7B & 2 & $1.6\times10^{19}$ & 2.8 & 24.3 / 51.0 & 1.91 & 0.09 (0.13) & 19.2 / 47.4 \\
\addlinespace
4B & 0 & -- & -- & 12.1 / 31.5 & -- & -- & 13.2 / 35.8 \\
4B & 2 & $4.0\times10^{19}$ & 6.1 & 37.9 / 65.6 & 3.21 & 0.07 (0.08) & 20.6 / 50.0 \\
\addlinespace
8B & 0 & -- & -- & 11.4 / 30.4 & -- & -- & 12.4 / 33.9 \\
8B & 2 & $7.7\times10^{19}$ & 8.5 & 49.2 / 74.6 & 4.18 & 0.04 (0.06) & 21.5 / 50.6 \\
\addlinespace
14B & 0 & -- & -- & 11.2 / 29.6 & -- & -- & 11.2 / 32.2 \\
14B & 2 & $1.5\times10^{20}$ & 14.4 & 64.3 / 84.8 & 5.26 & 0.03 (0.04) & \textbf{22.1} / \textbf{51.6} \\
\bottomrule
\end{tabular}

\end{table}

\subsection{Spread versus repeated exposures}
\label{app:spread}

The comparison of \cref{sec:spread} holds everything fixed except which queries a URL's samples carry. The 3,770,839 URLs of the 32M
subset with at least 16 training pairs, 98.8\% of them old pages, were trained from the same starting point with 16 samples each, placed in
one random order of sample slots that both arms share. Slot $k$ of a URL holds its $k$-th most distinctive pair in the spread arm and,
alternately, its primary or secondary pair in the repeated arm, whose two pairs are those of the grid. Each arm sees 60.3M samples with the
schedule of the grid, in about 50 minutes. The spread arm's samples are 3.6\% shorter in tokens, so its token-packed batches give 3,285
steps against 3,406 and 3.6\% less accumulated learning rate, which if anything favours the repeated arm.

We evaluate on the held-out queries whose target is among these URLs (2,597 of the 4,000) and on the pairs of 3,000 random URLs among
them whose query occurs only once among the spread arm's pairs (37,216, of which 5,590 are primary or secondary pairs; \cref{tab:s2div}).
Spreading the samples raises held-out hit@1 by $12.0\pm1.7$ points and hit@10 by $21.7\pm1.8$. It also ranks the 14 pairs that only the
spread arm trains first 22.0\% of the time, against 9.1\% for the repeated arm, which never saw these queries ($+12.9\pm0.4$), while the
two shared pairs are ranked first $54.8\pm1.3$ points less often. What repetition adds stays with the two trained queries: the repeated
arm leaves the likelihood of a URL's other queries unchanged at the first identifier level and 0.18 nats lower at the second.

We call a held-out query similar to a training query if they share at least half of their words (token Jaccard of at least 0.5) or
differ by edits of at most a quarter of the longer one. By this measure, 76.7\% of the held-out queries are similar to one of their URL's
16 training queries, and 35.3\% to one of the two shared ones. Spreading gains $11.5\pm3.3$ points on queries similar to a shared query,
$15.1\pm2.7$ on those similar only to a query that the spread arm alone trains, and $7.3\pm2.8$ on the 23.3\% similar to none (from 6.8\% to
14.0\%; 5.8\% before training). On the 8.8\% that share no word with any of the 16, the gain is $7.9\pm4.8$.

\begin{table}[H]
  \centering
  \scriptsize
  \caption{\textbf{Spread against repeated exposures} on the 3.8M URLs of the 32M subset with at least 16 training pairs: 16 samples per URL,
  either the grid's two pairs eight times each or the 16 most distinctive pairs once each, in the same sample positions. Shared pairs are
  the two pairs that both arms train (5,590 queries), other pairs the remaining 14, which only the spread arm trains (31,626); pairs
  ranked first per URL are estimated over all 16 from these two groups. Held-out queries are the 2,597 of the 4,000 whose target is
  among these URLs; similar ones (1,992) share at least half of their words with, or are within edits of a quarter of the length of, one
  of their URL's 16 training queries, dissimilar ones (605) with none.}
  \label{tab:s2div}
  \begin{tabular}{lccccc}
\toprule
Arm & Shared pairs & Other pairs & Pairs ranked first & Held-out queries & Held-out hit@1 \\
 & hit@1 / hit@10 & hit@1 / hit@10 & per URL (of 16) & hit@1 / hit@10 & similar / dissimilar \\
\midrule
Starting point (no samples) & 8.2 / 25.3 & 6.9 / 22.6 & 1.1 & 12.0 / 35.5 & 13.9 / 5.8 \\
Repeated: 2 pairs $\times$ 8 & \textbf{79.3} / \textbf{93.0} & 9.1 / 25.8 & 2.9 & 18.6 / 40.8 & 22.1 / 6.8 \\
Spread: 16 pairs $\times$ 1 & 24.5 / 53.8 & \textbf{22.0} / \textbf{51.7} & \textbf{3.6} & \textbf{30.6} / \textbf{62.5} & \textbf{35.6} / \textbf{14.0} \\
\bottomrule
\end{tabular}

\end{table}

\subsection{New pages}
\label{app:newpages}

Three measurements support \cref{sec:newpages}. First, how often a page was trained on explains most of the gap between old and new
pages. In the 4B model trained from its base checkpoint for 6,000 steps, pairs drawn once reach 44.2\% hit@1 on \trafficjul{} for old
pages and 10.1\% for new ones. Compared instead by how often the target URL was drawn as the target of any training sample, 0, 1 or 2--3
times, old pages reach 8.6\%, 12.4\% and 17.4\% hit@1 and new pages 6.8\%, 9.7\% and 12.2\% (29.7\%, 37.7\% and 44.8\% against 20.3\%,
27.1\% and 32.4\% at hit@10).

Second, document-indexing samples help new pages. They are given to a random subset of new pages in each round; in round 3, pages with
one reach 45.4\% hit@1 against 36.8\% for pages without, whereas the two groups were level at the end of round 2.

Third, a sample that carries only the URL teaches little. We tested it by repeating the 8M-URL grid run with its URLs split at random:
70\% trained on their pairs, 15\% on two exposures of a sample whose prompt is the normalised URL alone, and 15\% on nothing. On their test
pairs the three groups reach 67.6\%, 20.5\% and 17.2\% hit@1 for the 4B model (31.9\%, 20.5\% and 19.3\% for the 0.6B model), against
17.4--18.9\% for the starting point: the URL-only sample teaches 0.35 nats per pair at 4B against 3.93 for two query pairs (0.37 against
1.59 at 0.6B).

\end{document}